\documentclass[fleqn,usenatbib]{mnras}

\usepackage[symbol]{footmisc}
\usepackage{ae,aecompl}
\usepackage{graphicx}	
\usepackage{amsmath}	
\usepackage{amssymb}	
\usepackage{array}
\usepackage{longtable}
\usepackage{lscape}
\usepackage{dcolumn}
\usepackage{bm}
\usepackage{ifthen}
\usepackage{booktabs}
\usepackage{ulem}
\usepackage{geometry}
\usepackage{tikz}
\usepackage{rotating}
\usepackage[final]{pdfpages}
\usepackage[T1]{fontenc}
\usepackage{ae,aecompl}
\usepackage{rotating}
\usepackage{xcolor}

\newcommand{\himf}{H{\sc i}MF}

\def\H{H\,{\sc i}\,}
\def\b{BUDH{\sc i}ES\,}
\def\h{H{\sc i}MF\,}
\def\O{$\Omega_\mathrm{{H{\sc I}}}$}

\def\al1{ALFA100 }

\def\lgm{log($\mathrm{M_{H{\sc I}}}$/$\mathrm{M_{\odot}}$)}
\def\ms{$M_{H{\sc I}}^{\ast}$}

\def\ms{M$_{\odot}$}
\def\s{$\:$}

\def\kms{km s$^{-1}$}

\title[HI, optical and UV imaging of galaxies at z$\simeq$0.2]{BUDHIES IV: Deep 21-cm neutral Hydrogen, optical and UV imaging data of Abell 963 and Abell 2192 at z $\simeq$ 0.2 }

\author[Gogate et al.]{A.R. Gogate$^{1}$\thanks{E-mail: avanti@astro.rug.nl}, M.A.W. Verheijen$^{1}$, B.Z. Deshev$^{1, 3}$, J.H. van Gorkom$^{4}$,
\newauthor
M. Montero-Casta\~no$^{5}$, J.M. van der Hulst$^{1}$, Y.L. Jaff\'e$^{6}$, 
B.M. Poggianti$^{7}$ \\
$^{1}$ Kapteyn Astronomical Institute, University of Groningen, Landleven 12, 9747 AD Groningen, the Netherlands.\\
$^{2}$ Astronomical Institute, Czech Academy of Sciences, Bo\u{c}n\'i II 1401,
CZ-14131 Prague, Czech Republic \\
$^{3}$ Columbia University, 550 W 120th St, New York NY 10027, United States\\
$^{4}$ Dunlap Institute for Astronomy and Astrophysics, 50 St George St, Toronto, ON M5S 3H4, Canada \\
$^{5}$ Instituto de F\'isica y Astronom\'ia, Universidad de Valpara\'iso, Avda. Gran Breta\~na 1111, Casilla 5030, Valpara\'iso, Chile  \\
$^{6}$ INAF, Osservatorio Astronomico di Padova, Vicolo dell'Osservatorio 5, 35122, Padova, Italy\\}

\date{Accepted 2020, May 30. Received 2020 May 30; in original form 2019 December 30}

\pubyear{2019}

\begin{document}
\label{firstpage}
\pagerange{\pageref{firstpage}--\pageref{lastpage}}
\maketitle

\begin{abstract}
In this paper we present data from the Blind Ultra-Deep \H\s  Environmental Survey (\b), which is a blind 21-cm \H spectral line imaging survey undertaken with the Westerbork Synthesis Radio Telescope (WSRT). Two volumes were surveyed, each with a single pointing and covering a redshift range of 0.164 < z < 0.224. Within these two volumes, this survey targeted the clusters Abell 963 and Abell 2192, which are dynamically different and offer unique environments to study the process of galaxy evolution within clusters. With an integration time of 117$\times$12h on Abell 963 and 72$\times$12h on Abell 2192, a total of 166 galaxies were detected and imaged in \H. While the clusters themselves occupy only 4 per cent of the 73,400 Mpc$^3$ surveyed by \b, most of the volume consists of large scale structures in which the clusters are embedded, including foreground and background overdensities and voids. We present the data processing and source detection techniques and counterpart identification based on a wide-field optical imaging survey using the Isaac Newton Telescope (INT) and deep ultra-violet GALEX imaging. Finally, we present \H\s and optical catalogues of the detected sources as well as atlases of their global \H\s properties, which include integrated column density maps, position-velocity diagrams, global \H  profiles, and optical and UV images of the \H sources. 
\end{abstract}

\begin{keywords}
galaxies: evolution -- radio lines: galaxies -- galaxies: clusters: general -- galaxies: photometry
\end{keywords}



 \section{Introduction}\label{intro}

Fundamental properties of galaxies are shaped by internal processes during and after their formation (nature) but are also found to be strongly influenced by their environment (nurture). The $\Lambda$CDM cosmological framework suggests that galaxies form in dark matter structures and do not evolve in isolation \citep{Efstathiou90, Suginohara91, Gnedin96}. Within the $\Lambda$CDM model, many galaxies are predicted to transition from low to high-density environments at some point in their evolution. Galaxy properties such as star formation activity \citep{Balogh99, Poggianti06, DeLucia12}, morphology \citep{Dressler80} and gas content \citep{Denes14} have been found to have a strong dependence on the environment in which these galaxies reside. \citet{Dressler80} found high-density regions like clusters to be dominated by early type and lenticular galaxies, unlike the general field, which is mostly dominated by gas-rich spirals and dwarfs. Similarly, the Star Formation Rates (SFR) of galaxies are greatly reduced in the interiors of clusters as compared to their outskirts \citep[][and references therein]{Peng10}.

One of the main physical processes that cause environmentally driven evolution in and near galaxy clusters is ram pressure stripping \citep{Gunn72, Farouki80, Chung09, Oosterloo05, Poggianti17, Jaffe18}, a notion also supported by simulations \citep{Vollmer03, Tonnesen09, Kapferer09}. Other mechanisms, like tidal interactions and mergers \citep{Holmberg41, Toomre72, White78}, harassment \citep{Moore96, Smith10} and strangulation \citep{Larson80, Balogh00, Bekki02, Kawata08, Maier16} also play a significant role in the depletion of gas and the quenching of the SFR. These processes are also more dominant in high-density environments like groups and clusters and observational evidence is provided by studies such as those by \citet{Poggianti01} and \citet{Owers19}, pointing towards the truncation and eventual exhaustion of the atomic hydrogen (\H)  in galaxies as they approach the core of clusters. These accreted cluster galaxies may also have been members of smaller groups, making preprocessing an important factor in the $\Lambda$CDM structure formation scenario, as indicated in simulations \citep{Berrier09, McGee09, Han18}, as well as large optical surveys, such as the SDSS and 2dF surveys \citep{Lewis02, Gomez03}.

Due to their sensitivity to external perturbations and being the reservoir that fuels star formation, \H disks of galaxies prove to be ideal tracers for evolutionary processes. Blind \H imaging surveys have been crucial in the unbiased quantification of the environmental dependence of galaxy evolution \citep{Williams81, Chung09, Ramatsoku16}. While this is the case, \H  imaging requires long integration times which proves to be challenging at redshifts beyond 0.08. Consequently, very few blind \H surveys at higher redshifts, such as the Arecibo Ultra-Deep Survey \citep[AUDS,][]{Hoppmann15} and the COSMOS \H Large Extragalactic Survey \citep[CHILES,][]{Fernandez13, Hess19}, have been carried out so far. 

The cosmological evolution of the above mentioned astrophysical processes becomes evident at higher redshifts. For instance, \cite{BO84} (BO84 hereafter) found an unusually high fraction of blue, star-forming galaxies in cluster cores at higher redshifts compared to the present epoch. This trend, known as the "Butcher-Oemler Effect" (BO effect), was later confirmed by photometric as well as spectroscopic studies \citep[e.g.][]{Couch94,Couch98, Lubin96, Margoniner00, Lavery86, Couch87, Ellingson01, Tran03}. Numerous other studies have addressed the cause and nature of the BO effect \citep{Tran05, DePropris04, Urquhart10, Lerchester11}. Trends that are likely related to the BO effect were also found, for example, by \citet{Poggianti99}, \citet{Fasano00} and \citet{vanDokkum00} who found an increase in the spiral fraction in clusters at higher redshifts at the expense of the lenticular (S0) fraction \citep{Dressler97}, while studies by \citet{Duc02, Saintonge08} and \citet{Haines09} found an increase in the fraction of dusty blue galaxies in clusters with increasing redshift. The fraction of early-type galaxies in the field was also found to be reduced at higher redshifts \citep[e.g.][]{Bell07}. Based on these findings, some fundamental questions arise regarding the physical processes that govern galaxy evolution at higher redshifts, such as the effectiveness of ram-pressure stripping, the possibility of a higher gas content of infalling galaxies \citep{Catinella08}, different cluster accretion rates \citep{Berrier09}, or the possibility of blue cluster galaxies simply being post-starburst or backsplash systems that are no longer actively star forming \citep{Haines09, Oman13, Bahe13}. It is also argued that the BO effect is the result of an optical selection bias by preferential inclusion of clusters with bluer populations \citep[e.g.][]{Andreon04, Andreon06}.

While the BO effect has been extensively studied in the optical and infrared, there has been no study to date concerning the gas content of these blue galaxies in clusters at higher redshifts. With the idea of obtaining an \H  perspective on the nature of the blue galaxies responsible for the BO effect, we have conducted a deep, blind \H  imaging survey known as the Blind Ultra-Deep \H  Environmental Survey (\b) carried out with the Westerbork Synthesis Radio Telescope (WSRT). The main aim of this survey is to study two galaxy clusters, Abell 963 and Abell 2192  at z $ \simeq 0.2$ corresponding to a look-back time of $ \sim$ 2.5 Gyr, since signatures of cosmic galaxy evolution start becoming evident at such redshifts. Abell 963 is a BO cluster with a large fraction of blue galaxies in its core, while Abell 2192 has no identified population of blue galaxies associated with it but seems to host star forming galaxies in its outskirts based on the detection of O\textsc{ii} emission lines by \citet{Yara0_12}. The properties of the two clusters are described in detail in Sect. \ref{targets}. 

Apart from the \H  data presented in this paper, ancillary optical imaging data is available in  $B-$ and $R-$ bands obtained with the Isaac Newton Telescope (INT) and presented in this paper. In addition, deep imaging data is available at ultra-violet wavelengths (NUV and FUV) obtained with $GALEX$, as well as optical photometry and spectroscopy provided by the SDSS and supplemented with optical spectroscopy obtained with the William Herschel Telescope \citep[WHT,][]{Yara1_13}. Targeted CO observations of 23 galaxies within the two surveyed volumes has also been obtained with the Large Millimeter Telescope (LMT) \citep{Cybulski16}. Additionally, 89 and 111 redshifts were also available in Abell 963 from Lavery \& Henry \citep{Lavery86} and Czoske (private communication) respectively. Furthermore, optical redshifts are available for Abell 963 from \citet{Hwang14} and from the Local Cluster Substructure Survey \citep[see][]{Yara3_16}.

A \b pilot study of these two clusters with integration times of 20 $\times$ 12$^{hr}$ for Abell 963 and 15 $\times$ 12$^{hr}$ for Abell 2192 was presented by \citet{Verheijen07}. Upon stacking of the \H spectra, they found no evidence of \H in the cluster core of Abell 963. These observations, however, were of limited sensitivity and revealed only a fraction of \H detections obtained from the full survey. A total of 42 galaxies were detected in \H that had optical counterparts in SDSS, and the lowest detected \H mass was 5 $\times 10^9$ M$\odot$. The full survey, details of which are provided in Sect. \ref{obs}, is much more sensitive and provides a better picture on the gas content in the cores of the two clusters and other overdensities in the surveyed volume. A detailed environment characterization by \citet{Yara1_13}, based on optical redshifts, showed significant substructure associated with both clusters. For Abell 963, \citet{Yara2_15, Yara3_16} constructed a phase-space diagram of the cluster galaxies and carried out some preliminary \H  stacking. They deduced that the large fraction of blue galaxies observed in the core of Abell 963 may be the result of preprocessing, having caused a temporary enhancement of star formation, while the galaxies lost their gas down to the \b detection limit during their first infall due to ram-pressure stripping. 

\begin{figure*}
\includegraphics[scale=0.47]{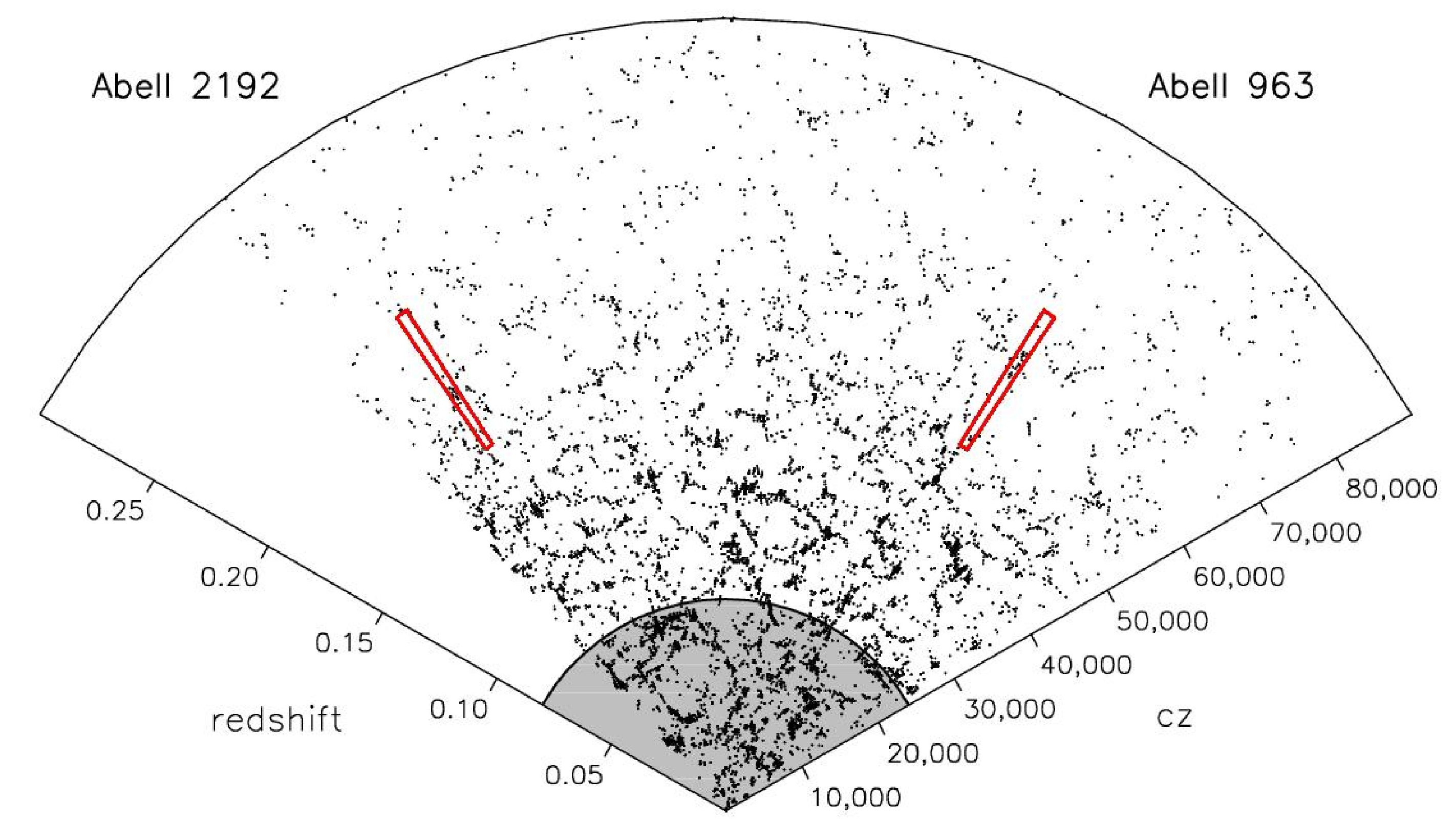}
\caption{An SDSS pie diagram including both clusters taken from \citet{Verheijen07}, showing the distribution of SDSS galaxies out to z $\approx$ 0.3. The grey shaded area indicates the extent of the ALFALFA and HIPASS surveys. The two red boxes show the two volumes surveyed by \b, demonstrating that not only the clusters but also large volumes in front and behind the clusters are included, encompassing all cosmic environments.}
\label{fig:sdsspie} 
\end{figure*}

In this paper, along with the Westerbork, INT and GALEX data, including source catalogues and an atlas, we present a preliminary analysis of the \H, optical and UV data of the galaxies within the surveyed volume in preparation of a more detailed comparative study of the two clusters with a focus on the BO effect. Additionally, the \b data will be used in a forthcoming paper to measure the \H Mass Function (\h) and the cosmic \H density (\O), from direct \H detections at z $\simeq$ 0.2 by virtue of its total survey volume of 73,400 Mpc$^3$, encompassing a wide range of cosmic environments. The \h and \O\s have been well constrained in the Local Universe \citep{Rosenberg02, Zwaan05, Springob05, Martin10, Jones18} but only one study exists based on direct \H detections at a higher redshift \citep{Hoppmann15} out to $z \approx 0.16$. Another application of the data will be the study of the \H-based Tully-Fisher relation \citep[TFr,][]{Tully_Fisher77} at z $\simeq$ 0.2. The \H-based TFr, though extensively used in the Local Universe, cannot be easily studied at higher redshifts \citep{Catinella15} by cause of the intrinsic weakness of the \H signal. Other gas tracers like CO are therefore used beyond the Local Universe \citep[e.g.][]{Topal18}, though the CO emission does not usually extend beyond the peak of the rotation curve and into the Dark Matter halo. Bright emission lines from the ionized ISM, such as H\textsc{$\alpha$}, H\textsc{$\beta$},  O\textsc{ii} and O\textsc{iii} are also often used \citep[e.g.][]{Flores06, Kassin07, Puech08, DiTeodoro16}. Like CO, however, their presence is confined to the inner part of the rotation curve since they emit within the stellar disk. The choice of tracer, therefore, may lead to systematic differences in the kinematic measures and hence in the statistical properties of the TFr \citep[][and references therein]{deblok14,deblok16}. With our \H data, we have identified several TFr candidate galaxies for a forthcoming study of the \H-based TFr at z $\approx$ 0.2. This will allow us to directly and consistently compare the \H-based TFr obtained from Local Universe observations with our work at a higher redshift.

\begin{table}
\begin{center}
    
\begin{tabular}{ccc}
\hline
\hline
                                                        & A963                   & A2192             \\ \hline 
RA                                                      & 10h17m14.22s      & 16h26m36.99 s \\
Dec                                                     & +39d01m22.1s           & +42d40m10.1 s \\
z                                                       & 0.206$^{(1)}$       & 0.188$^{(1)}$ \\
Richness class                                          & 3$^{(2)}$                      & 1$^{(2)}$  \\
L$_x$                                                   & 3.4 $\times$ 10$^{44}$$^{(3)}$ & 7 $\times$ 10$^{43}$$^{(4)}$ \\
$\sigma$                                                & 993$^{(5)}$                   & 653$^{(6)}$               \\
n$_{det}$ & 127                    & 39 \\
f$_{B}$ &19$^{(7)}$  & $-$ \\
\hline              
\end{tabular}
\end{center}

\caption{General properties of the two clusters. Top to bottom: Pointing centre coordinates [J2000], Redshift of the cluster, richness class, X-ray luminosity [ergs s$^{-1}$], velocity dispersion [km s$^{-1}$], number of \H detections and the fraction of blue galaxies [\%].\newline [1] \citet{Yara1_13}, [2] \citet{Abell58},[3] \citet{Haines18},[4] \citet{Voges99}, [5] \citet{Yara3_16}, [6] \citet{Yara0_12}, [7] \citet{BO84}}
\label{tblprop}
\end{table}

This paper is organized as follows: In Sect. \ref{targets} we provide a detailed description of the two target volumes. Sect. \ref{obs} summarizes the WSRT observing strategy  and includes technical details of the observations, data processing, source finding, completeness tests as well as an introduction to the \H\s catalogues, samples of which are given in tables \ref{tab:A963_HI_table} and \ref{tab:A2192_HI_table} for A963 and A2192 respectively. Details on the INT wide field optical imaging are provided in Sect. \ref{wf}, as well as information on the optical source catalogues, given in tables  \ref{tab:A963_optical_table} and \ref{tab:A2192_optical_table}. The full \H\s and optical catalogues are provided as supplementary online material. Sect. \ref{uv} presents the UV imaging data and the procedures undertaken for the data reduction and source finding. In Sect. \ref{results} we discuss the observed and derived \H\s properties of the galaxies, show the colour-magnitude diagram and present the \H\s atlas for all the \H\s detected galaxies. A brief analysis of the data sets is summarized in Sect. \ref{summ}.  Throughout this paper, we assume a $\Lambda$CDM cosmology, with $\mathrm{\Omega_{M}}$ = 0.3, $\Omega_{\Lambda}$ = 0.7 and a Hubble constant H$_0 = $70 km s$^{-1}$ Mpc$^{-1}$.

\section{Targets}\label{targets}
\b is a blind 21 cm survey comprising two single-pointing fields, each containing an Abell cluster along with the large scale structure surrounding them. The two volumes encompass a wide range of environments, which includes the two clusters along with smaller groups, sheets and large voids. The two volumes are indicated by the two red boxes in the pie-diagram from the SDSS footprint in Fig. \ref{fig:sdsspie}. These two clusters, Abell 2192 at z $\simeq$ 0.188 and Abell 963 z $\simeq$ 0.206, occupy only $\sim$ 4\% of the total surveyed volume. They were chosen to represent a well studied BO cluster (Abell 963) and a control cluster (Abell 2192), both very distinct in their dynamical properties. The clusters and the volumes containing them will henceforth be referred to as A963 and A2192 respectively. The properties of both clusters are summarized in Table \ref{tblprop}. 
A963, which is also in the seminal BO84 sample, is a massive, virialised lensing BO cluster, bright in X-rays \citep{Allen03, Smith05}. It was chosen as one of the two \b\s targets because of its unusually large fraction of blue galaxies (19\%); the largest blue fraction at z $\sim$ 0.2 in the BO84 sample. It contains a cD galaxy with a stellar mass of $10^{12}\: \mathrm{M{_\odot}}$, surrounded by multiple blue arcs of lensed star-forming background galaxies at $0.731 < z <3.269$ \citep{HL84}. The estimated total mass of the cluster is 1.4 $\times$ 10$^{15}$ M$_{\odot}$ \citep{Yara3_16}. \citet[][see Fig. 6]{Haines18} found three X-ray groups falling into the main cluster. While the core of A963 is relaxed, the outskirts show a large degree of substructure, according to the environmental analysis undertaken by \citet{Yara1_13}, who found two groups within the cluster and two structures outside the cluster (see Fig. 10 in their paper).  In the foreground, they found an overdensity in the same field-of-view which is group/sheet-like and separated from A963 by a large void. Another overdensity is located in the background of A963, well outside its turnaround radius, and does not show much evidence for substructure. In A963, a total of 134 galaxies with optical redshifts were identified in the magnitude range $m_3$ and $m_3 + 2$, where $m_3$ is the magnitude of the third brightest cluster member, corresponding to a cluster richness class 3 \citep{Abell58}.

Similar to A963, the volume containing A2192 also consists of a range of environments, which includes three distinct overdensities and two voids. \citet{Yara1_13} found that the central overdensity comprises the cluster A2192, which is dynamically younger and less massive than A963. With a total mass estimated at 2.3 $\times$ 10$^{14}$ $h^{-1}$ M$_{\odot}$, A2192 has a large degree of substructure and is in the process of accreting a nearby compact group and a population of gas-rich, field-like galaxies \citep{Yara0_12}. It is very weak in X-rays \citep{Voges99}, and the blue fraction for this cluster is still unknown. In the foreground and background of A2192, there exist group-like overdensities separated from the cluster by large voids. A2192 is of richness class 1 and 62 cluster galaxies were identified in the magnitude range $m_3$ and $m_3 + 2$ \citep{Abell58}.

\section{Westerbork observations and data processing} \label{obs}

The \H\s imaging survey was carried out with the Westerbork Synthesis Radio Telescope (WSRT) during eight semesters between 2005 and 2008, using the cooled Multi-Frequency Front Ends and the digital DZB backend. The 14 dishes of the WSRT provided baselines ranging from 36 to 2700m. To reach a minimum detectable \H\s mass of $2\times 10^9$ M$_\odot$ at the field centres at their respective cluster redshifts over an emission line width of $\sim$150 \kms and a signal-to-noise of 4 in each of three adjacent spectral resolution elements, A963 was observed for a total of 118$\times$12$^{\rm hr}$ and A2192 for a total of 72$\times$12$^{\rm hr}$ (see Table \ref{tab:nrobs}). The integration time was 60 seconds as a compromise between tangential smearing near the edges of the field and a manageable data volume.  The primary beam Full Width at Quarter Maximum (FWQM) is 61 arcminutes at 1190 MHz or corresponding to z=0.194 for \H\s emission. The observations of A963 were centred on 4C +39.29 ($\alpha$=10:17:14.20, $\delta$=+39:01:21.6, J2000), a 1.13 Jy bright radio continuum source at 2.4 arcminutes to the south-east of the cluster core. The observations of A2192 were centred on the cluster proper at ($\alpha$=16:26:37.00, $\delta$=+42:40:10.8, J2000). To obtain complex gain and bandpass calibrations, each 12-hour measurement was preceded and followed by a 30 minute observation of a flux calibrator: 3C147 and CTD93 for A963, and 3C286 and 3C48 for A2192.

The backend and correlator were configured to cover 1160$-$1220 MHz in dual-polarisation mode with eight, partially overlapping IF bands each 10 MHz wide and divided into 256 channels, providing a channel width
of 39.0625 kHz corresponding to a rest-frame velocity width of 9.84 km/s at 1190 MHz. The spectra were Hanning smoothed to suppress the Gibbs phenomenon near the bandpass edges, resulting in a velocity resolution of 19.7 km/s. This setup remained unchanged throughout all observations. For the \H emission line it covers a redshift range of 0.16427 < z < 0.22449, corresponding to a recession velocity range of 49,246 < cz < 67,300 kms$^{-1}$. Given the adopted cosmology, this redshift range corresponds to a luminosity distance of 789 < D$_{\rm lum}$ < 1117 Mpc, a range in look-back time of 2.05 < T$_{\rm lookback}$ < 2.69 Gyr, and a range in primary beam diameter of 7.09 < FWQM < 9.55 Mpc. The spatial scales at the distances of the two clusters are 3.4 kpc arcsec$^{-1}$ for A963 and 3.1 kpc arcsec$^{-1}$ for A2192. The total surveyed comoving volume within the FWQM of the primary beam is 73,400 Mpc$^3$, equivalent to a spherical volume of the Local Universe within 26 Mpc.

\begin{table}
\centering
\begin{tabular}{ccc}

\hline \hline
Volume & Year & Number of 12$^{hr}$ observations \\ \hline
A963   & 2005 & 23                     \\
       & 2006 & 34                     \\
       & 2007 & 31                     \\
       & 2008 & 30                     \\ \hline
A2192  & 2005 & 15                     \\
       & 2006 & 20                     \\
       & 2007 & 14                     \\
       & 2008 & 23                     \\ \hline
\end{tabular}
\caption{The number of 12$\mathrm{^{hr}}$ measurements obtained for both the \b volumes between 2005 - 2008.}
\label{tab:nrobs}
\end{table}

\subsection{Data flagging and calibration}\label{flg}

The visibility data obtained for \b\s were processed and Fourier transformed with the help of the NRAO Astronomical Image Processing System (AIPS) \citep{Greisen90}. The datacubes were further processed
and analyzed with the Groningen Image Processing SYstem (GIPSY) \citep{vdHulst92}.

\begin{figure}
\includegraphics[trim={7cm 0cm 7.8cm 0cm},clip, width=0.95\linewidth]{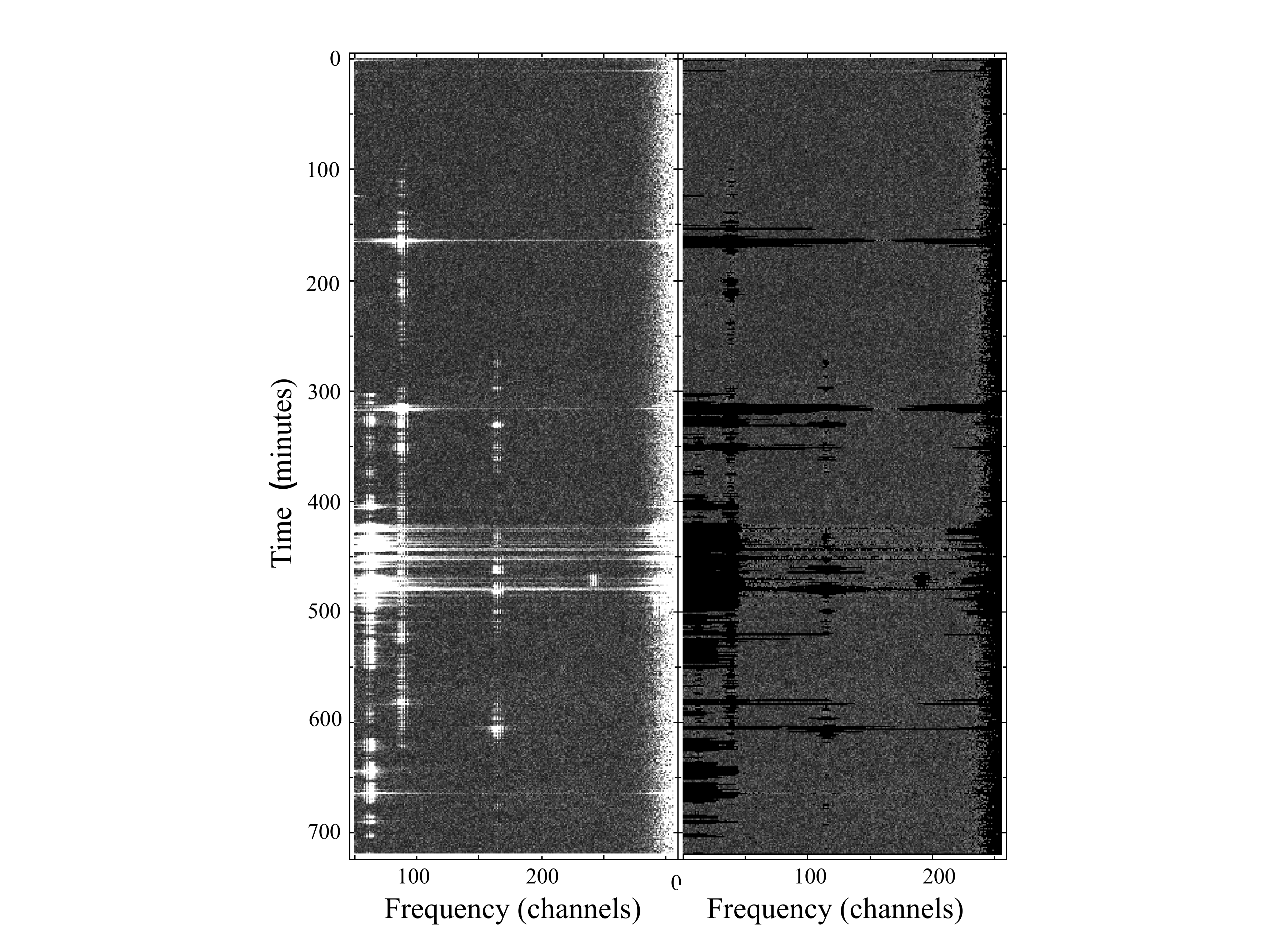}
\caption{An example illustrating the RFI removal algorithm adopted in the data processing for one polarisation of one IF of a random 12-hour measurement consisting of all 256 channels. The left and right panels show the data before and after RFI flagging. }
\label{fig:rfi}
\end{figure}

\begin{figure*}
\includegraphics[trim={1.5cm 0.3cm 1.7cm 1cm},clip,width=\textwidth]{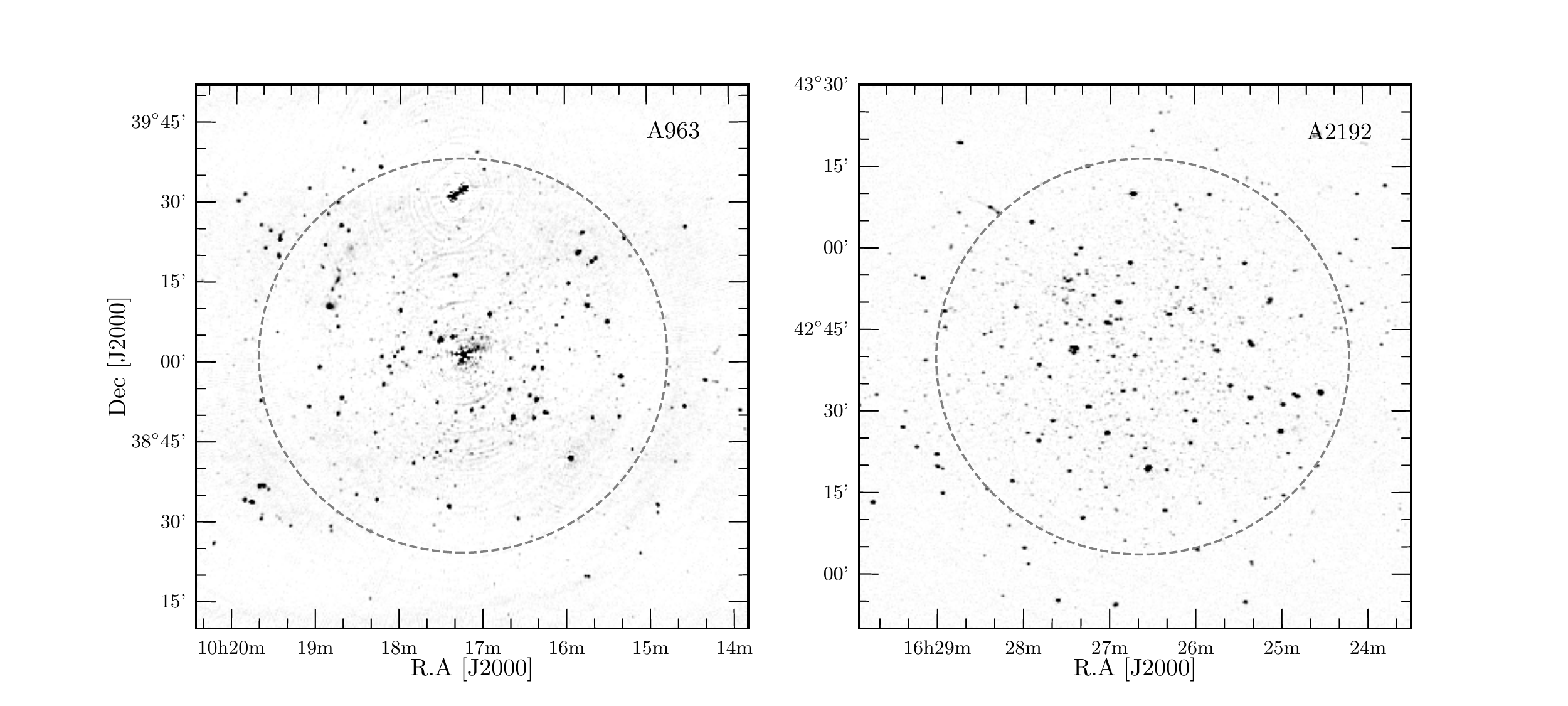}
\caption{Radio continuum maps of the two \b fields with self-calibration applied. The dashed grey circles indicate the FWQM of the primary beam at the redshift of the respective clusters.  The arcs observed in A963 are instrumental artefacts caused by the strongest continuum sources in the field, locally limiting the dynamic range of the image, and are completely unrelated to the physical lensing properties of A963.}
\label{fig:contmaps}
\end{figure*}

\begin{figure}
\includegraphics[width=1\linewidth]{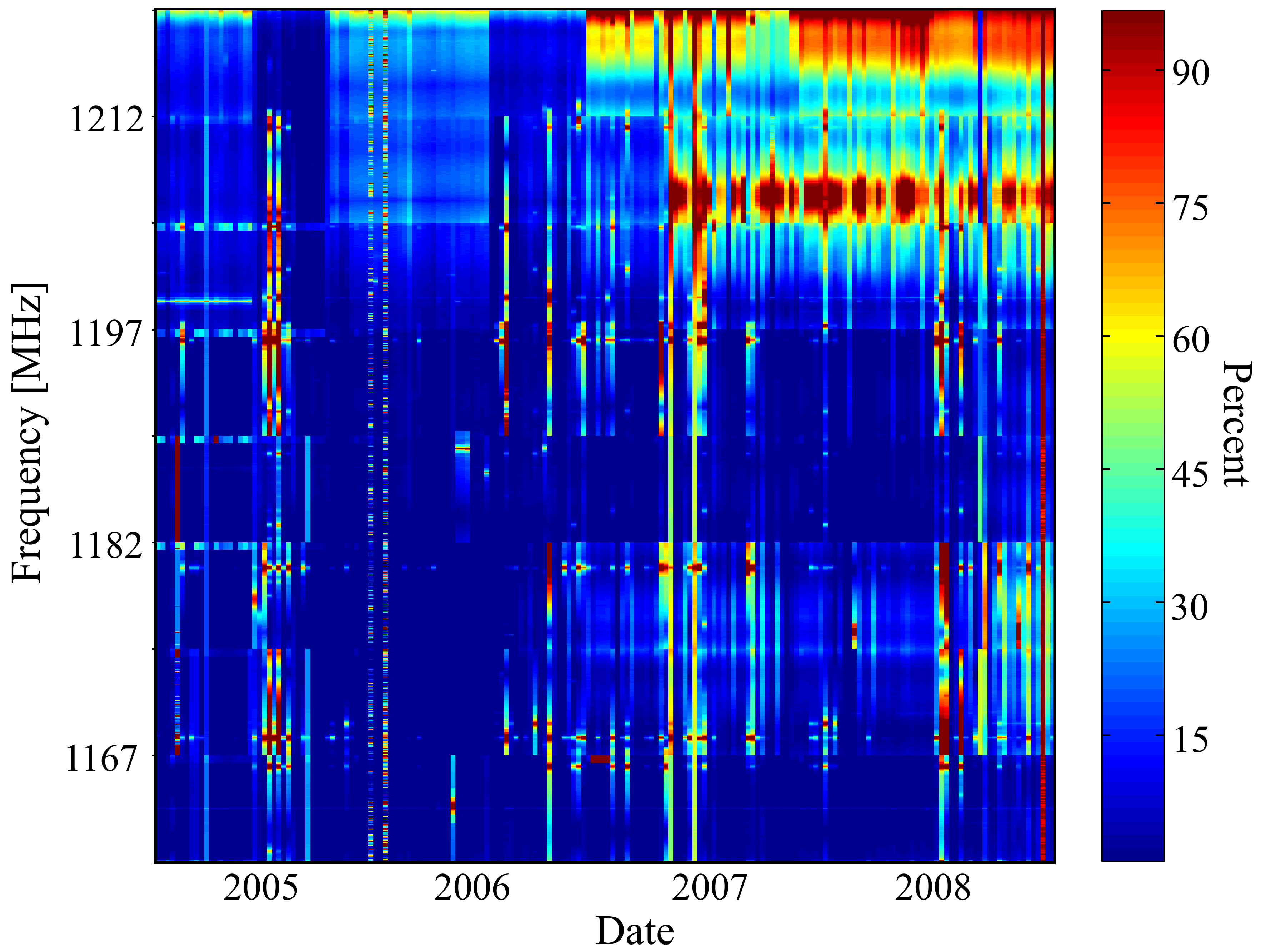}
\caption{RFI at different frequencies plotted as a function of time for all measurements combined. The colour range indicates the percentage of RFI-induced data loss.}
\label{fig:rficol}
\end{figure}

\subsubsection{Cross-calibration and data flagging}

The visibility data of each 12-hour measurement and the bracketing two calibrators were imported and combined into a single AIPS dataset.  The temporal behaviour of the system temperatures was inspected to assess the overall health of each antenna throughout the observation, and an initial flagging of disfunctional antennae was carried out manually.  Visibilities of the calibrators that were affected by strong Radio Frequency Interference (RFI) were flagged by clipping visibilities with excessive amplitudes.  Subsequently, a preliminary, normalised bandpass was determined for each of the two calibrators.  After applying this antenna-based bandpass, a linear fit was made to the amplitudes of each baseline-based spectrum and visibilities with an amplitude in excess of 8$\sigma$ above this fit were flagged.  After this flagging, a new, normalised bandpass was determined for each calibrator, each polarisation and each of the eight IF bands separately. 

\noindent 
After applying the new bandpass to the calibrator data, the central 75 percent of each 10 MHz IF band was averaged to form continuum datasets for the calibrators.  The observed, complex continuum visibilities were then calibrated to match the expected, known flux density of the calibrators.  These complex gain and bandpass solutions were applied to the spectral line visibilities of the science fields by interpolating the solutions for the two calibrators in time across the 12-hour measurement.  This was done for each of the eight IF bands independently. 

\noindent
After this cross-calibration, RFI was removed from the science data in an iterative manner.  A linear fit was made to the visibility amplitudes of the central channels (26 to 217) and subtracted.  Each visibility spectrum was then boxcar-smoothed with different kernel widths to reduce the noise.  After each smoothing operation, visibilities with an amplitude in excess of 4.5$\sigma$ were flagged, followed by a new linear, censored fit.  After several iterations, the accumulated flags were applied to the original, cross-calibrated visibility data of the science targets. Figure \ref{fig:rfi} shows the results of this RFI removal algorithm.

\subsubsection{Self-calibration  and Continuum subtraction}\label{selfcal}

After cross-calibration and flagging, the visibilities of each measurement of the target fields were self-calibrated for each IF separately by iteratively building a sky model of the continuum sources in each field, consisting of clean components.  For each IF, the central channels were averaged to produce a continuum data set, which was Fourier transformed to the image domain and cleaned with the standard H\"ogbom algorithm, using manually controlled search areas centred on visually identified continuum sources out to a distance of several degrees from the phase centre. The clean components were subsequently used to derive and apply corrected gain solutions over increasingly shorter time intervals, followed by a new Fourier transform to produce an improved image.  For the field containing A963, four phase-only self-calibrations were carried out with solution intervals of 10, 5, 2 and 1 minute, followed by three phase-and-amplitude calibrations with solution intervals of 15, 10 and 5 minutes, for each IF separately.  For the field with A2192, the four phase-only self-calibrations were sufficient.  The self-calibration results were inspected visually at every step.  This resulted in sixteen sky models, one for each of the eight IFs for each of the two fields. The accumulated gain corrections were applied to the corresponding spectral line data and the clean components that produced the final self-calibration solutions were subtracted from the visibilities using \textit{uvsub} followed by \textit{uvlin} in order to remove the continuum flux from the spectral line data.

\noindent
Figure \ref{fig:contmaps} shows the continuum maps for both fields. These maps were made separately at a higher resolution than the line data, using most of the available 60 MHz band. Use of a {\it Robust} weighting of -5 (uniform weighting) resulted in Gaussian beams with Full Width Half Maxima (FWHM) equivalent to 15.7" $\times$ 23.2" for A963 and 20" $\times$ 20" for A2192.  The noise is not uniform over the entire image because source confusion and dynamic range artefacts locally enhance the noise in the field centres above the thermal level, making the estimation of the rms noise in the maps problematic. In the corners of the fields we measure average thermal noise values of 8 $\mu$Jy/beam for A963 and 6.5 $\mu$Jy/beam for A2192. Figure 3 of \citet{Zwart15} provides a theoretical confusion noise limit at 1.4 GHz as a function of angular resolution. For our synthesized beam sizes, our measured thermal noise values are below the expected confusion noise of $\sim$13 and $\sim$14 $\mu$Jy/beam for A963 and A2192 respectively, at 1.19 GHz.

\subsection{Imaging data cubes}\label{dc}

The self-calibrated and continuum subtracted visibility data of each IF of each measurement were Fourier transformed with \textit{imagr} with a {\it Robust} weighting parameter of +1 and no further baseline tapering.  This produced 'dirty' image cubes of 512$\times$512 pixels and 256 channels, as well as cubes with the frequency dependent synthesized beam pattern to be used for cleaning the \H\s line emission.  With 8$"$ pixels the channel maps covered 68$\times$68 arcmin$^2$ on the sky. The FWHM of the Gaussian beams with which clean components were restored are 23$"$ $\times$ 37$"$ for A963 and 23$"$ $\times$ 39$"$ for A2192 independent of frequency.  Data cubes from all epochs were averaged channel-by-channel with weights based on the measured rms noise in a channel.  The same weights were used to average the cubes with the synthesized beams.

\subsection{Residual Continuum subtraction}\label{contsub}
The combined image cubes, by virtue of their significantly lower noise, again revealed residuals from the brightest continuum sources, mainly due to temporal bandpass variations. The grating rings of the residuals were cleaned and the continuum sources themselves were masked out. The area removed from the survey due to these masks was < 2\% for the field of A963 and 0.5\% for A2192.

Fitting and subtraction of the continuum with \textit{uvlin} also removed some of the underlying \H signal, resulting in negative baselines in spectra containing \H emission. This bias was corrected in two steps. The first involved an iterative procedure in which the spectrum containing the \H signal was clipped, fitted and subtracted with the help of the GIPSY task \textit{conrem}. This step was repeated until all of the negative baselines were corrected for to the best possible extent. This step, however, may have clipped some of the low column density \H signal as well, which had to be restored. For this purpose, a source detection algorithm (see Sect. \ref{sf} for details on the algorithm) was run on the cubes which masked the \H signal recovered in the previous step. These masks were then applied to the cubes that still contained the imperfections due to \textit{uvlin}. A direct, censored \textit{conrem} was then used again, and the baselines were then fit linearly and subtracted, thus bringing all the negative baselines as close to zero as possible, without losing any \H signal due to clipping. The final cubes were free of offending imperfections, only containing \H signal.

\subsection{Quality of the data}\label{qdat}
\subsubsection{RFI}\label{RFI}
Figure \ref{fig:rficol} shows the trend in RFI at all frequencies as a function of time. The colours indicate the percentage of data lost due to RFI within the duration of the survey, and does not include data from antennae that did not work. From the figure, it is evident that the data are relatively RFI-free in the initial measurements between 2005 and 2006, with a drastic increase in RFI at higher frequencies in the more recent measurements post 2006. This increasing RFI is mostly caused by the buildup of navigation satellite constellations, particularly Galileo and GLONASS, which transmit in this frequency range.

\begin{figure}
\includegraphics[width=\linewidth]{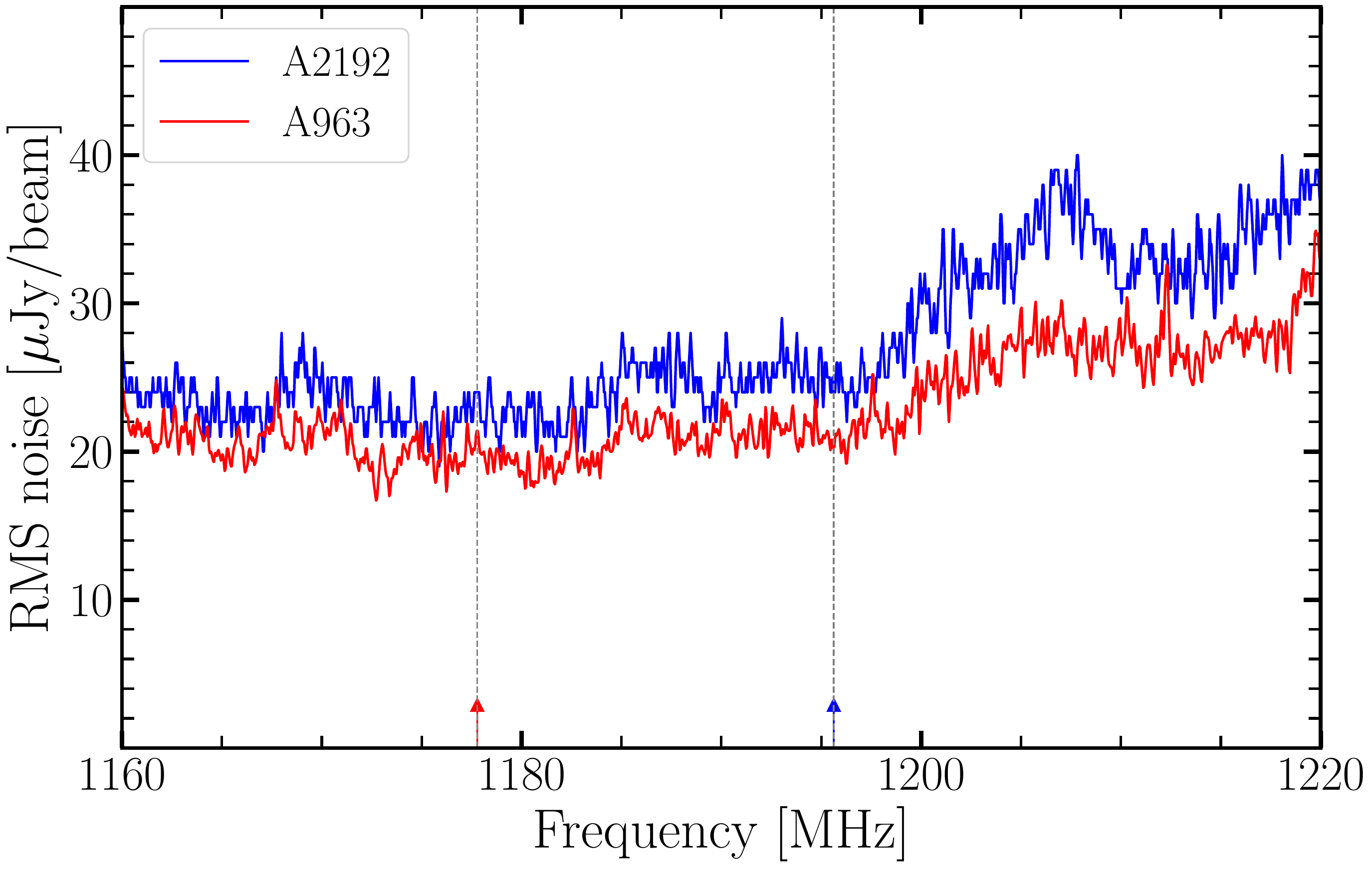}
\caption{The rms noise variation as a function of frequency in the continuum subtracted cubes for A963 (red) and A2192 (blue) when smoothed to a velocity resolution of 38 km s$^{-1}$. The trend with frequency is similar for the two cubes, and the peak in the rms noise is likely caused by low-level RFI at those frequencies, which could not be entirely removed from individual measurements. The vertical dashed lines with the colour-coded arrows correspond to the redshifts of the two clusters.} 
\label{fig:noise}
\end{figure}

\noindent
\subsubsection{Noise variation}
Figure \ref{fig:noise} shows the rms noise variation as a function of frequency in the two cubes after continuum subtraction and at a velocity resolution of 38 km s$^{-1}$ (R4, see Sect. \ref{sf}). The variation in both the volumes show very similar trends even though the data processing was carried out independently. The increase in noise, particularly towards higher frequencies above 1200 MHz can be attributed to the increase in RFI and frequency dependent flagging, with about 5 - 8 \% of the visibilities below 1200 MHz needed to be flagged, while more than 15 \% were flagged above 1200 MHz.  This increase in RFI is also seen in Fig. \ref{fig:rficol}, as explained in Sect. \ref{RFI}.  With achieved noise levels comparable to the expected thermal noise of the system, imperfections in the bandpass calibration become more apparent in the combined measurements, particularly the frequency-dependent residuals from those continuum sources which could not be properly subtracted. However, while one can avoid confusion with real sources due to the coherent spatial and frequency information that these residuals possess, they still add to the overall noise in channel maps.
The errors in the cubes are based on the rms noise only, and do not include uncertainties due to the masks or continuum subtraction imperfections.

\begin{figure}
\includegraphics[width=\linewidth]{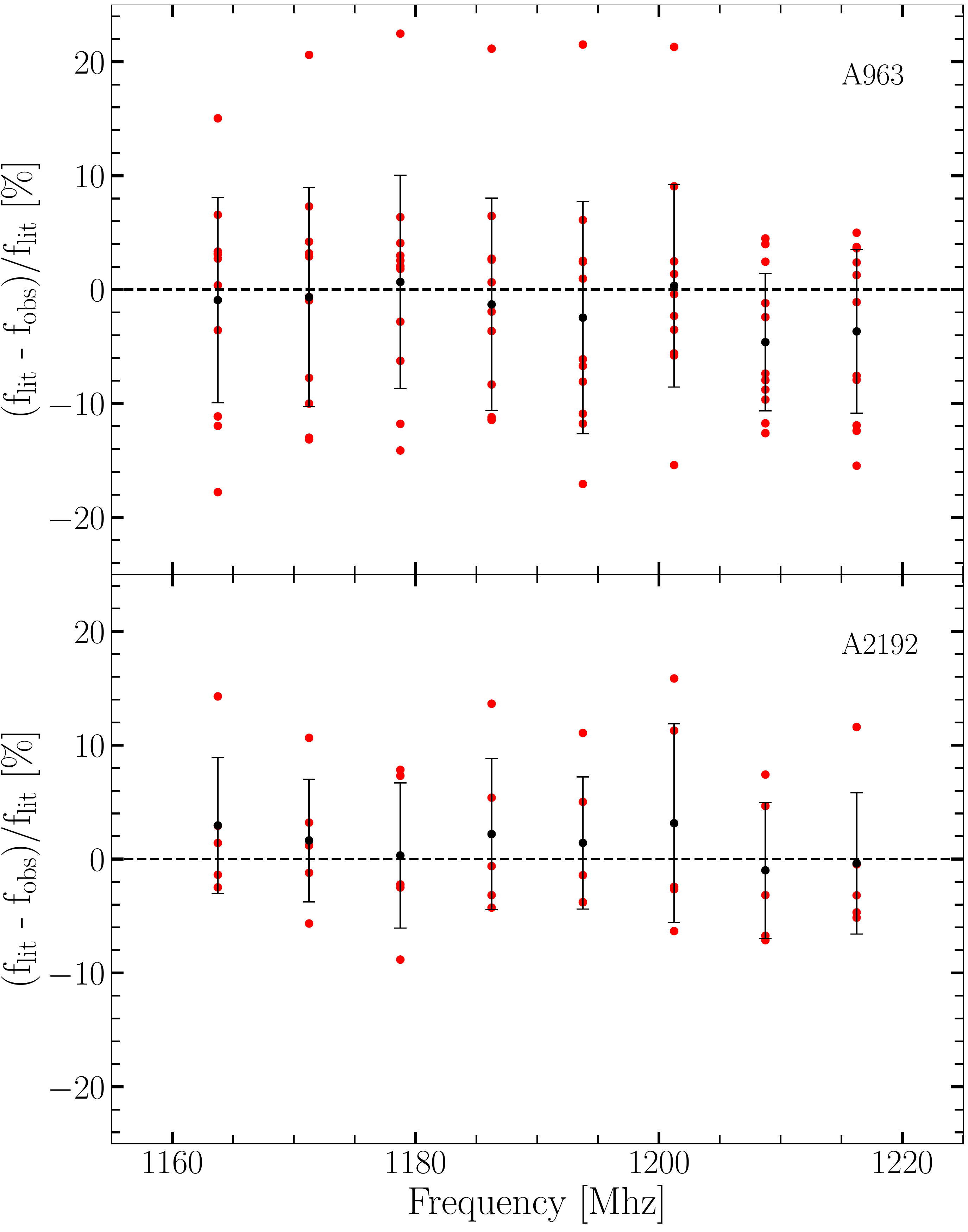}
\caption{Comparison of continuum flux densities measured with \b\s with those from FIRST and WENSS, shown as the mean flux difference between literature flux densities and those measured by us as a function of frequency (MHz). Top: A963, Bottom: A2192. Eleven continuum sources were chosen in the field of A963 and five in A2192. These sources are given by the red points. The expected fluxes for the eight IF bands of our measurements were calculated assuming a single spectral index between the known FIRST and WENSS fluxes. The black points and the error bars indicate the mean and standard deviation of the flux differences respectively at each IF frequency.}
\label{fig:cflx}
\end{figure}
\subsubsection{Comparison of continuum flux densities}
Since none of the \H\s sources were previously detected, we compared instead the flux densities of the detected continuum sources against those published in the literature, namely the FIRST \citep{Becker95} and WENSS \citep{WENSS97} surveys, which were used for their continuum flux measurements at 1400 and 325 MHz respectively. Eleven continuum sources in the field of A963 and five in that of A2192 were found to have reliable measurements in all the 8 IF bands of our survey, and also had both, FIRST and WENSS flux densities. Upon calculating the spectral index $\alpha$, from the two literature surveys, the expected flux density for a continuum source at each IF frequency from our survey was calculated. The differences between the expected and measured flux densities are shown in Fig. \ref{fig:cflx}, illustrating the mean flux difference of these continuum sources as a function of frequency. These differences in all bands were found to be consistently less than 10\%, within the errors. Outlying sources could be variable over time. We concluded that the continuum flux density measurements in all bands were consistent with those obtained by FIRST and WENSS. By inference, the \H\s fluxes we measure also have estimated calibration errors of $<$ 10\%.

\subsection{HI source finding and galaxy identification}\label{sf}
\begin{figure}
\includegraphics[width=\linewidth]{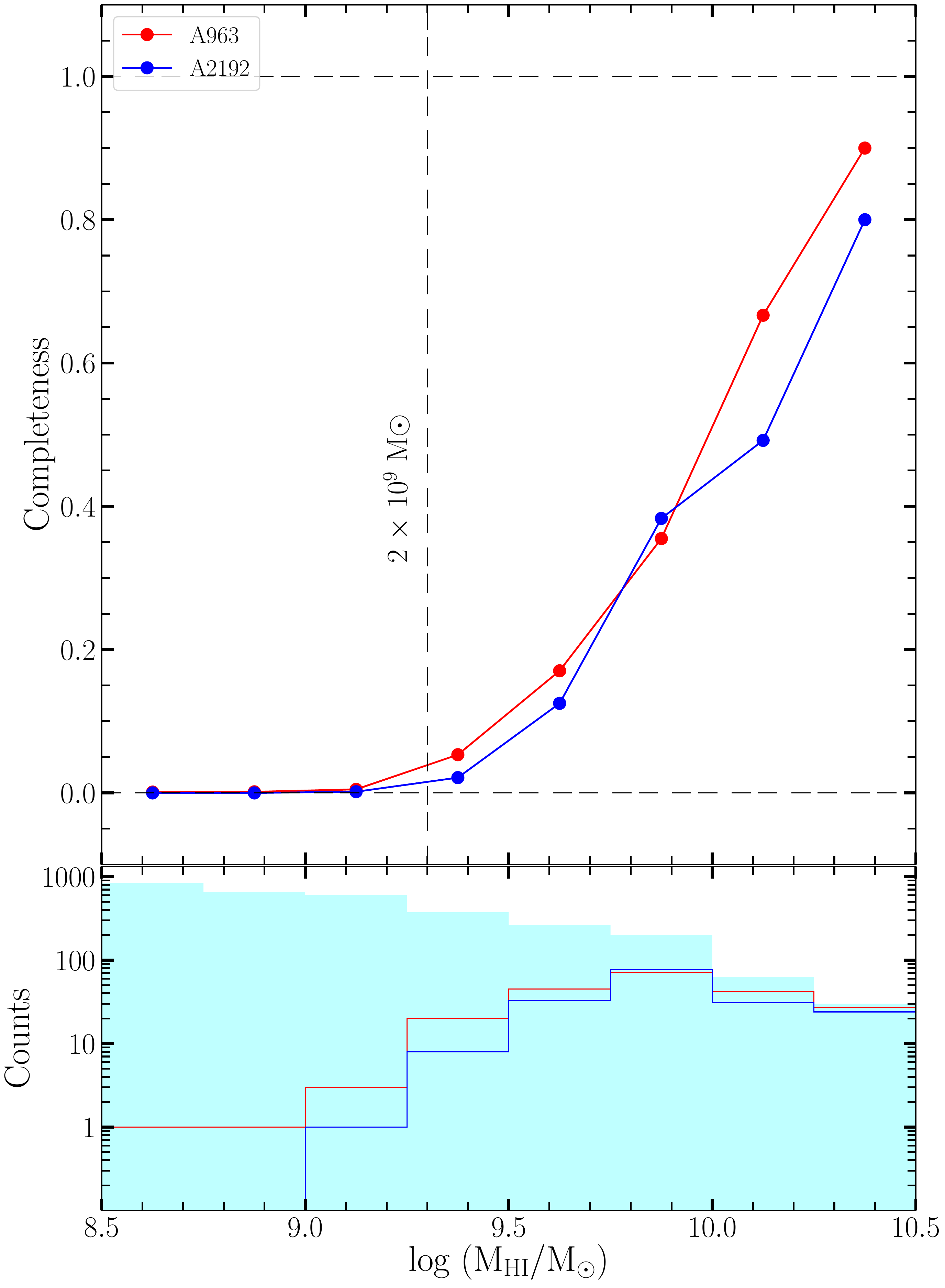}
\caption{ Top panel:} Completeness of the two surveyed volumes per \H\s mass bin. Red: A963; Blue: A2192.  Horizontal black dashed lines at the bottom and the top indicate  0 and 100 \% completeness  respectively, while the vertical dashed line indicates the intended detection limit of 2 $\times$ 10$^9$ M$\odot$ at the redshift of the two clusters. Bottom panel: injection and recovery rates of artificial galaxies in the two volumes. The inserted artificial galaxies are given by the cyan filled histogram, while the line histograms show the recovered galaxies in the two volumes, colour-coded as in the top panel.
\label{fig:compl}
\end{figure}

Source finding was carried out using GIPSY. The two processed cubes were first smoothed in velocity to four velocity resolutions consisting of an initial Hanning smoothing followed by further smoothing with a Gaussian kernel to a nearly Gaussian frequency response with a FWHM of 4, 6 and 8 channels. These four cubes, referred to as R2, R4, R6 and R8 hereafter, were made for each field. R2 corresponds to a Hanning smoothed cube (19 \kms), whereas the other resolutions correspond to Gaussian response functions corresponding to four (38 \kms), six (57 \kms) and eight (76 \kms) channels respectively at 1190 MHz. Locations in the cubes that contained residuals from strong radio continuum sources, as well as a perimeter of 5 pixels along the edges where aliasing effects occur, were masked out.  The pixels betweeen $\pm$ 3,4,5 and 8 times the rms noise in each channel were then clipped, and the remaining positive and negative pixels connected in multiple adjacent velocity channels were searched for. Negative pixels were included to estimate false detections. The detection criteria were selected to combat imaging artefacts and noise in the cubes. They are as follows: A single spectral resolution element with an 8$\sigma$ peak, two adjacent elements at 5$\sigma$, three adjacent elements at 4$\sigma$ and four adjacent elements at 3$\sigma$   were considered as solid detections, while the rest of the data was discarded. The algorithm returned three-dimensional masks of the detected sources. 
The source finding algorithm revealed 153 positive detections in the field of A963 and 41 in A2192. To confirm the reliability of our \H\s detections, these positive detections returned by the source finding process were corroborated  by the existence of optical and UV counterparts, located within the FWHM of the synthesized beam centred on the \H\s detection. We cross-matched our \H\s data with the SDSS, and our own wide-field optical images from the INT (see Sect. \ref{wf}) and ultra-violet images from GALEX (see Sect. \ref{uv}). In most cases, there is an obvious stellar counterpart within the HI contours. In many cases this galaxy also had a corresponding optical redshift, which unambiguously settled the identification.  From those unambiguous cases we learned that the corresponding galaxies were also bright in UV.  Therefore, if multiple, possible stellar counterparts existed within the \H\s contours and no optical redshifts were available to confirm or reject a counterpart, then we used the GALEX images and assigned the brightest UV source with an optical counterpart to be the plausible stellar counterpart of the \H\s detection. A total of 28 spurious \H\s detections without an optical or UV counterpart were rejected. The \H\s centres of all the confirmed \H\s detections were determined by fitting 2D Gaussians to the total \H\s maps. The final sample comprises 127 galaxies in A963 and 39 galaxies in A2192. These numbers are an update from those mentioned in previous \b\s publications, and are the result of a re-analysis of the line cubes which included the deblending of confused sources. Images of these optical and UV counterparts with overlaid \H\s contours are provided in the \H\s atlas (Sect. \ref{atlas}).

\begin{figure*}
\includegraphics[width=0.32\linewidth]{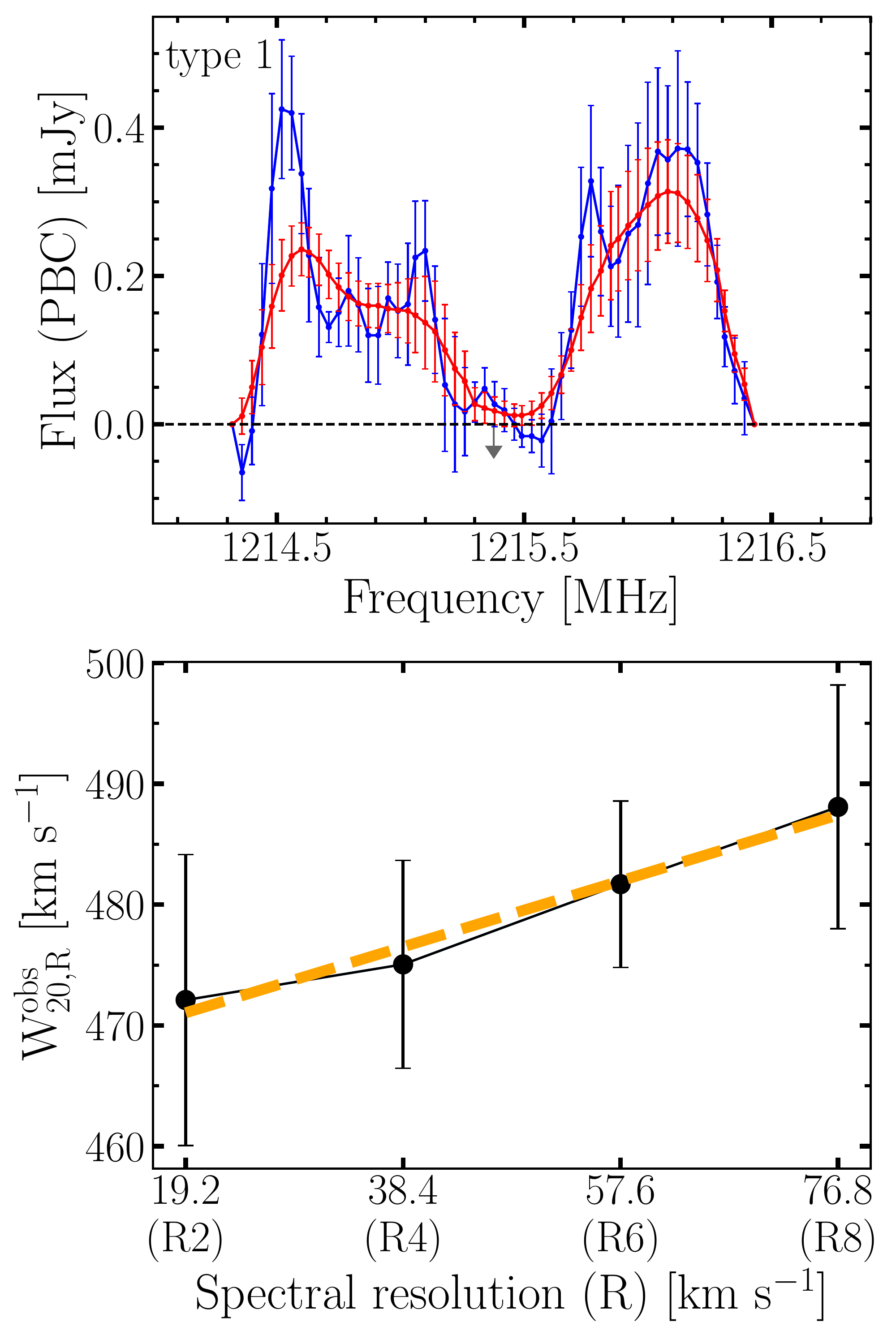}
\includegraphics[width=0.32\linewidth]{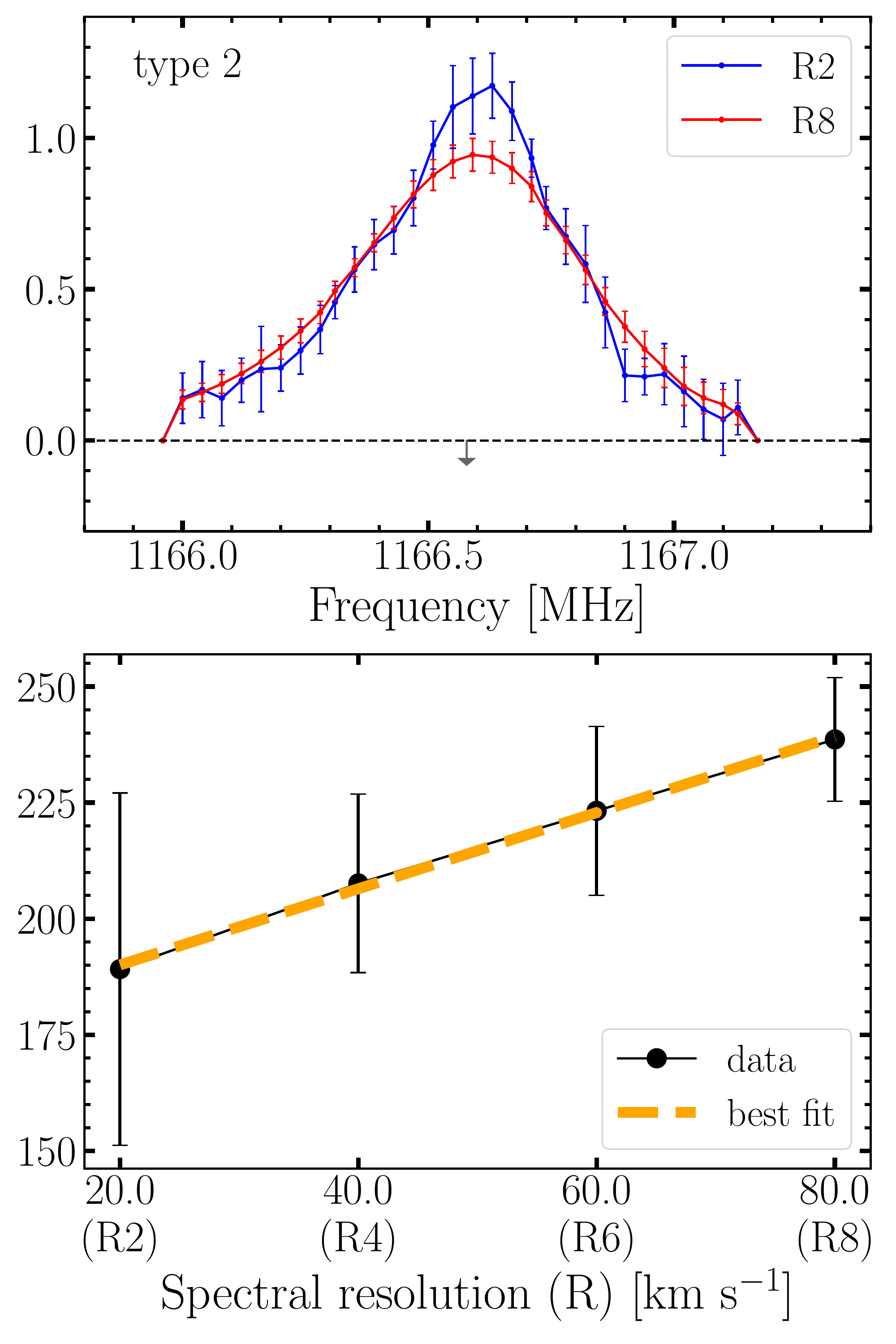}
\includegraphics[width=0.32\linewidth]{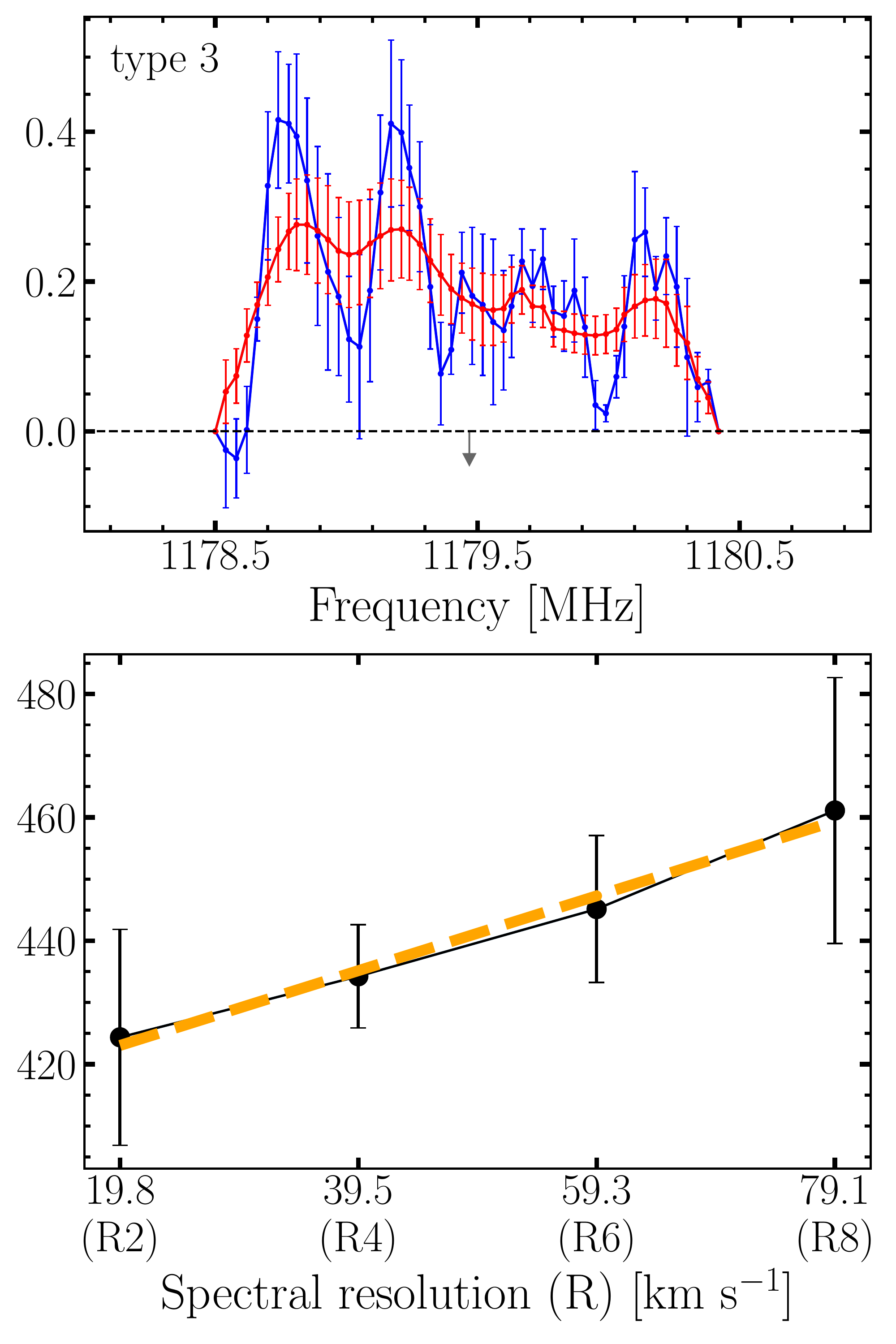}

\caption{Three galaxies chosen to represent the entire \b\s sample, showing the effect of instrumental broadening as a function of spectral resolution. The top panels show the global \H profiles at the R2 and R8 resolutions (red and blue lines respectively) along with errors. The grey arrows indicate their central frequency corresponding to their systemic velocities. For the entire sample, these profiles are broadly categorized as double-horned (type 1), single gaussian (type 2) and asymmetric (type 3). In the bottom panels we show in black the measured line widths against their respective resolutions: R2, R4, R6 and R8 (See sect. \ref{sf}), measured in the galaxy rest frame, along with errors. The dashed orange lines indicate the best linear fits. }
\label{fig:w20prof}
\end{figure*}

\subsection{Completeness}\label{compl}
We studied the completeness of the two detected galaxy samples based on an empirical approach by inserting artificial galaxies throughout the entire survey volume and subsequently determining the rate of recovery of these sources using the same source finding scheme as described in Sect. \ref{sf}. For the purpose of the completeness tests, a library of artificial galaxies in the mass range $10^{8.5}$ to $10^{10.5}$ M$_\odot$ was created following standard scaling relations (e.g. the \himf, the \H\s mass-diameter and the Tully-Fisher relation) and covered all inclinations. These galaxies were created using the GIPSY task \textit{galmod}. \\
A total of 3000 galaxies were inserted into two synthetic cubes corresponding to the two volumes such that they followed the observed cosmic large scale distribution of galaxies in these volumes. 700 of these galaxies were above the nominal survey detection limit of 2 $\times 10^9$ \ms. These noise-free synthetic cubes were then multiplied by the primary beam and added to the observed data cubes. The new data sets containing real and artificial sources were searched again for \H detections using the same source detection method. To reduce confusion, the real sources recovered in the new data cubes were first subtracted before identifying the artificial galaxies. Source identification was carried out by cross-matching the positions of the recovered artificial sources with the input catalogue. Some sources needed to be manually identified due to the effects of blending.  The search recovered a total of 210 artificial sources for the volume of A963 and 169 sources for A2192. The completeness of the survey was estimated as the ratio of  input to recovered artificial sources per mass bin. This is illustrated in Fig. \ref{fig:compl}, from which it is evident that the survey is not fully complete in any of the mass bins, with a rapid decrease in the recovery of galaxies towards lower masses. The detectability of galaxies depends on three main parameters: firstly their position in the cube, and their distance and thereby the extent of the attenuation by the primary beam; secondly, their inclinations, and thirdly, the non-uniform noise distribution. Consequently, we found that at higher redshifts and galaxies near the edges of the field of view, as well as those that are highly inclined have a lower detection probability. The process of source insertion, detection and identification carried out for the completeness tests will be described in more detail in an upcoming publication focusing, for the first time, on the \h\s and \O\s based on direct \H\s detections at z $\sim$ 0.2. 

\subsection{HI properties of detected galaxies}\label{gloprop}
This section describes the methods involved in measuring the \H\s properties of the individual galaxies detected in both volumes. Some of these properties are included in the \H source catalogues and atlas, described separately in Sect. \ref{cat} and Sect. \ref{atlas} respectively. 

\subsubsection{Global \H\s profiles}\label{gp}
The primary beam corrected fluxes required for the global \H profiles were determined using the GIPSY task \textit{flux} within the frequency-dependent \H\s emission masks made during source finding and applied to the R2 cubes. The uncertainty in the flux calculation was determined by first applying the \H\s mask to 8 line-free areas in the cube over the same channels as the actual line, offset by one synthesised beam from the source and then subsequently measuring the signal in these offset regions. The rms scatter in these 8 measurements provides an empirical uncertainty in the flux values obtained for each galaxy. Peak flux densities were then measured for three equal velocity sections of the profiles, corresponding to the approaching (F$\mathrm{^{max}_{app}}$), middle (F$\mathrm{^{max}_{mid}}$) and receding (F$\mathrm{^{max}_{rec}}$) bins, and were used in quantifying the shape of the global profile: double-horned (type 1), single Gaussian-like (type 2) or asymmetric (type 3) respectively. These three types of profiles are illustrated in Fig. \ref{fig:w20prof}.

\begin{figure}
\includegraphics[trim={1cm 0.2cm 1cm 0.2cm},clip,width=1\linewidth]{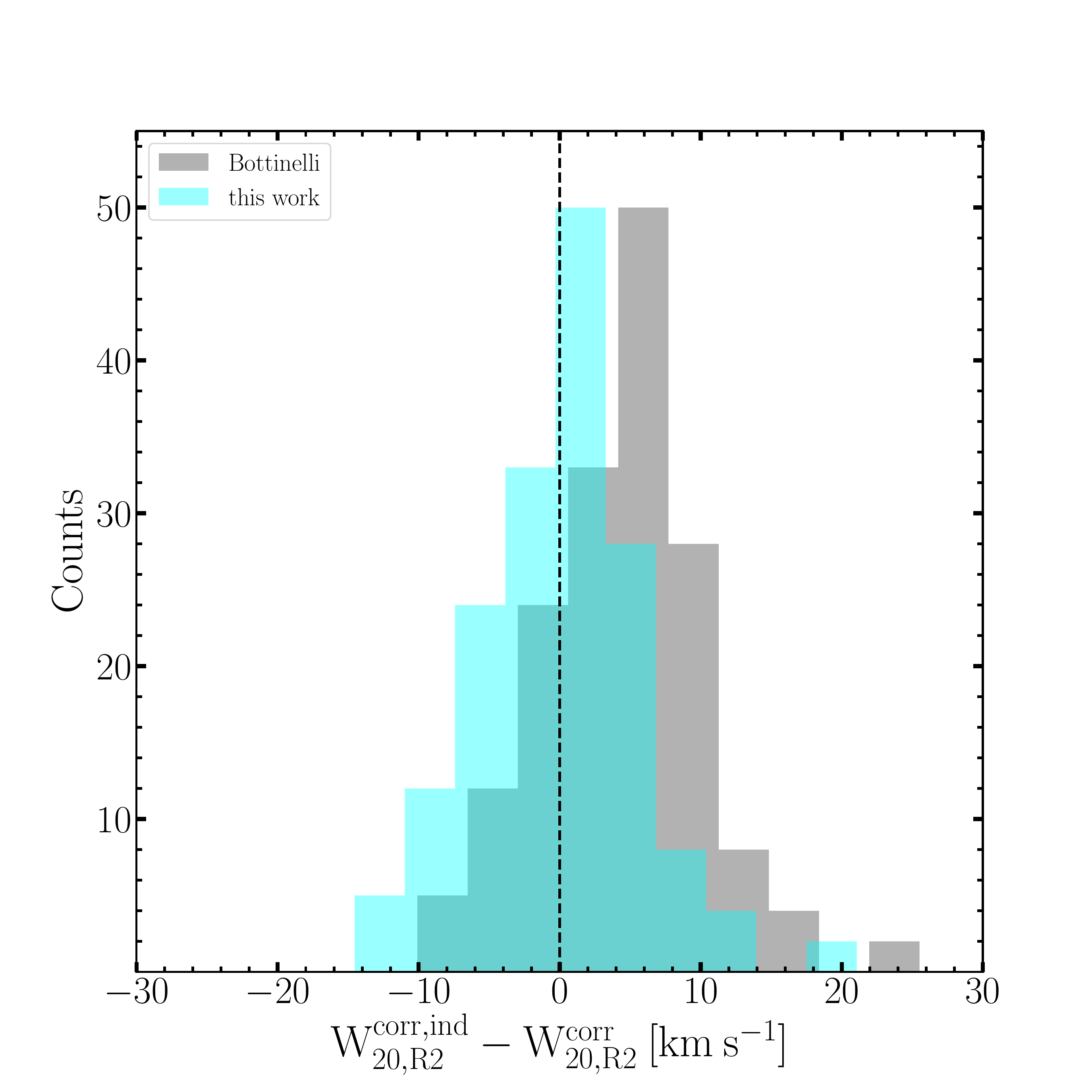}
\caption{Histograms of the difference between the expected and corrected W$_{20}$ line widths at the velocity resolution R2. The cyan histogram represents corrected W$_{20}$ widths based on the difference between Eq. \ref{eq:w20corr} and Eq. \ref{eq:wcorrind}. Given as reference is a grey histogram that follows a similar procedure but is based on the correction factor taken from \citet{Bottinelli90}. The overlap between the two histograms is shown in dark cyan, and the dashed line implies zero difference between the expected and corrected W$_{20}$ widths at R2.} 
\label{fig:w20hist}
\end{figure}

The observed frequency widths [MHz] at 20 \% and 50 \% ($\Delta \nu_{obs,\%}$) of the overall peak flux densities were determined at each resolution (R) based on the widths obtained from a linear interpolation between the data points moving outwards from the profile centre on both the approaching and receding sides ($\nu_{r,\%}$ and $\nu_{a,\%}$ respectively). These observed line widths were converted to velocity widths [km s$^{-1}$] as per Eq. \ref{eq:lw1}.

\begin{equation}
\mathrm{W^{obs}_{\%,R}} = \frac{\mathrm{\Delta \nu_{obs,\%}}}{\mathrm{\nu_{rest}}}(1+z)c
\label{eq:lw1}
\end{equation}

\noindent
where $\mathrm{\nu_{rest}}$ is the rest frequency of the Hydrogen emission line.
The errors on the line widths were calculated from the errors on the global profiles, also based on Eq. \ref{eq:lw1}. 

\noindent
The rest-frame channel widths in km s$^{-1}$ were calculated as 

\begin{equation}
\mathrm{\Delta V^{chan}_{rest} = c \frac{\Delta f} {\nu_{obs}}}
\label{eq:dVrest}
\end{equation}

\noindent
where $\mathrm{\Delta f}$ is the width of the spectral channel in MHz and $\nu_{\mathrm{obs}}$ is the observed frequency of the \H\s line. The total primary beam corrected \H\s flux (S) for each galaxy was calculated by integrating their global profiles, given by 

\begin{equation}
\mathrm{S^{V_{rest}}  = \Sigma S_\nu \Delta V^{chan}_{rest}}
\label{eq:totflx}
\end{equation}

\noindent
where S$\mathrm{^{V_{rest}}}$ is in Jy km s$^{-1}$ and S$_\nu$ is the primary beam corrected flux density within the mask in each channel, given in Jy.

The observed line widths (Eqs. \ref{eq:lw1}) of galaxies increase with decreasing spectral resolution, and hence need to be corrected for instrumental broadening. Ideally, the corrected widths at all velocity resolutions should be the same, or, in other words, a line fit to the corrected widths plotted as a function of velocity resolution, should have a zero slope. 
\noindent
For each source, a first-order polynomial was fit to the observed rest frame line widths in km s$^{-1}$ at each of their respective four resolutions R2, R4, R6 and R8  (see Sect. \ref{sf}). An average correction factor C was then derived from all the slopes of the individual fits. Under the assumption that the required correction factor (C) is linear with resolution (R), the universal correction ($\mathrm{\delta W_{\%,R}}$) to be applied to the measured line widths of galaxies at different resolutions was calculated as:
\begin{equation}
\delta \mathrm{W_{\%,R} = C \times R}
\label{eq:corrfac}
\end{equation}

\noindent
The observed line widths were then corrected following: 
\begin{equation}
     \mathrm{W^{corr}_{\%} = W^{obs}_{\%,R} - \delta W_{\%,R}
\label{eq:wcorr20}}
\end{equation}

\noindent
where $\mathrm{W^{corr}_{\%}}$ is the corrected line width after applying the correction factor to the observed line width $\mathrm{W^{obs}_{\%,R}}$. Figure \ref{fig:w20prof} illustrates three galaxies chosen to represent the entire sample. The global \H profiles in the top panels are shown for two resolutions, R2 (black) and R8 (blue). One can notice that the line width increases at lower spectral resolutions. In the bottom panels we plot the observed line widths as a function of restframe velocity resolution, with the dashed line indicating the line that best fits the data. The corrections in the line widths were calculated based on Eq. \ref{eq:corrfac}. The final corrected line widths in km s$^{-1}$ are calculated according to:
\begin{equation}
    \mathrm{W_{20}^{corr}} = \mathrm{W_{20,R}^{obs} - 0.36R}
\label{eq:w20corr}
\end{equation}

\begin{equation}
     \mathrm{W_{50}^{corr}} = \mathrm{W_{50,R}^{obs} - 0.29R} 
\label{eq:w50corr}
\end{equation}

\noindent
To confirm that these corrected line widths were not systematically over/under-corrected, we first calculated the expected linewidth per galaxy based on the individual correction factor (c$\mathrm{^{ind}}$) obtained from the slopes of the respective polynomial fits, such that Eq. \ref{eq:corrfac} becomes

\begin{equation}
 \delta \mathrm{W_{\%,R}^{ind} = c^{ind} \times R}
\end{equation}

\noindent
and Eq. \ref{eq:wcorr20} becomes

\begin{equation}
   \mathrm{W^{corr,ind}_{\%} = W^{obs}_{\%,R} - \delta W^{ind}_{\%,R}} 
     \label{eq:wcorrind}
\end{equation}

\noindent
where $\delta\mathrm{W\mathrm{_{\%,R}^{ind}}}$ is unique for each galaxy and would be the ideal correction required in order to have equal line widths at all resolutions. For each galaxy, the corrected line widths from Eq. \ref{eq:wcorr20} were subsequently subtracted from these expected line widths (Eq. \ref{eq:wcorrind}). Ideally, this difference should amount to zero for all galaxies.
We compared these values with corrected W${\mathrm{^{corr}_{\%,R}}}$ line widths based on \citet{Bottinelli90}, who claim a correction factor of 0.55R and 0.19R for W$\mathrm{^{obs}_{20,R}}$ and W$\mathrm{^{obs}_{50,R}}$ respectively. The histograms in Fig. \ref{fig:w20hist} illustrate this comparison of corrected W$\mathrm{^{corr}_{20,R}}$ line widths at the R2 resolution. From the figure, it is clear that our corrections have a closer convergence around zero compared to those from \citet{Bottinelli90}, which are slightly offset. We find similar results for the W$\mathrm{^{corr}_{50,R}}$ line widths.

\begin{figure*}
\includegraphics[ width=0.95\linewidth]{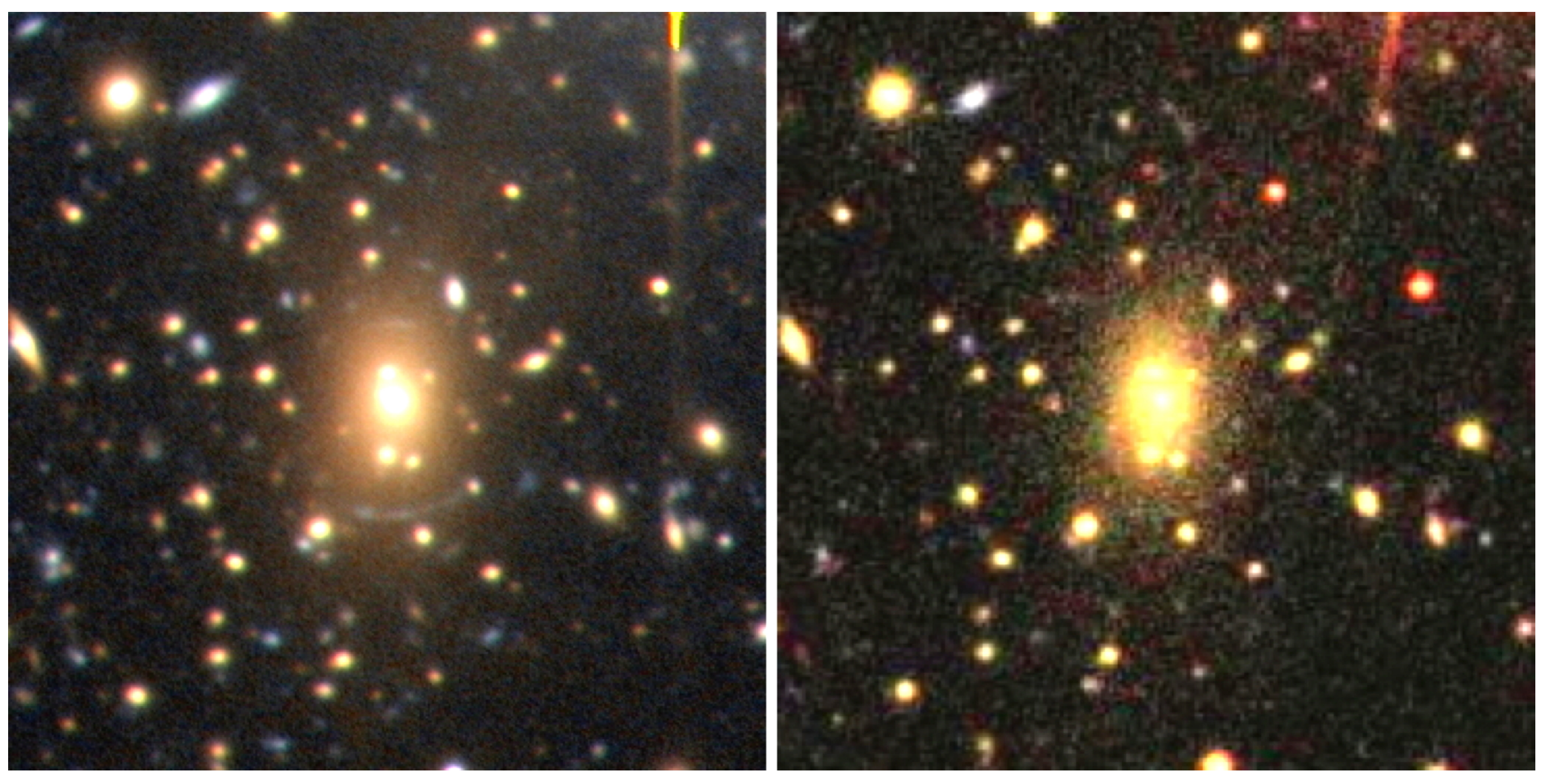}
\caption{A comparison of the area within the central 2$\times$2 arcmin$^2$ region of A963, showing our composite optical INT  image (left) and SDSS (right). The contrasts in both images have been increased to demonstrate the depth. The INT images were created using Harris B- and R- bands, while the SDSS image is a composite of g, r and i bands. Multiple arcs and low surface brightness sources can be observed around the central cD galaxy in our image. Note the number of blue galaxies in the centre of A963.}
\label{fig:sdssvsint}
\end{figure*}

\subsubsection{\H\s masses and column density maps}\label{mom0}
The total \H\s  mass M$\mathrm{_{HI}}$ (M$_\odot$) of each galaxy was calculated as:
\begin{eqnarray}
\mathrm{M_{HI} \: = \frac{2.36 \times 10^5}{(1+z)} \: \times D_{lum}^2 \: S^{V_{rest}}}
\label{eq:mhi}
\end{eqnarray}

\noindent
where D$\mathrm{_{lum}}$ (Mpc) is the luminosity distance to the galaxy, based on its redshift and the adopted cosmology. \\

Total \H maps were constructed from the emission within masks that outlined the \H\s emission from the cleaned data cubes. All the channels were then co-added to make the maps, which were subsequently corrected for primary beam attenuation. Though most galaxies in our volumes were barely resolved, we computed column densities from the \H\s maps by converting pixel values from Jy beam$^{-1}$ to cm$^{-2}$ by using Equations \ref{eq:nhi} and \ref{eq:tb}. These column densities were used for choosing appropriate \H\s contour levels (see Sect. \ref{atlas}). 
\begin{eqnarray}
\mathrm{N_{HI} = 1.82 \times 10^{18} \int T_{b}\: dV_{rest}^{chan}}
\label{eq:nhi}
\end{eqnarray}

\noindent
where T$\mathrm{_{b}}$ is the brightness temperature, given by

\begin{eqnarray}
\mathrm{T_b \: [K] = \frac{6.05 \times 10^5 }{\Theta_x \Theta_y} (1 + z)^3 \: S_\nu} \:[\mathrm{Jy \: beam^{-1}}]
\label{eq:tb}
\end{eqnarray}

\noindent
Here, $ \mathrm{\Theta_x }$ and $\mathrm{\Theta_y}$ correspond to the major and minor axes of the clean beam in arcseconds, and z is the redshift of the \H\s emission line. It is to be noted, that total \H\s column density maps are meaningful only for spatially resolved sources. The equations \ref{eq:lw1}, \ref{eq:dVrest}, \ref{eq:totflx}, \ref{eq:mhi}, \ref{eq:nhi} and \ref{eq:tb} have been adopted from \citet{Meyer17} and take cosmological effects into account.

For extended sources larger than the synthesized beam, the column density sensitivity at 5 times the rms noise ($\mathrm{N_{HI}^{5\sigma}}$)
 at the redshift of the clusters is 0.91$\times$10$^{19}$ cm$^{-2}$ for A963 and 1.1$\times$ 10$^{19}$cm$^{-2}$ for A2192. The rms values are based on Fig. \ref{fig:noise}. These column density limits are remarkably low and can be attributed to the exceptionally long integration times.

\subsubsection{Position-Velocity diagrams}
Position-velocity diagrams (PVD) along the kinematic major axis give an impression of the projected rotation curve of the \H\s disc of a galaxy. For our sample, PVDs were extracted from the cubes with velocity resolutions of 19 km s$^{-1}$ (R2) and 76 km s$^{-1}$ (R8) respectively. The centres of the PVDs are based on the \H centres, while the position angles were obtained from SExtractor fits made to the R-band INT data as described in Sect. \ref{wf}.

\subsection{The HI catalogue}\label{cat}
Tables \ref{tab:A963_HI_table} and \ref{tab:A2192_HI_table} list the derived \H\s parameters for all confirmed \H\s detected galaxies in the two volumes. The full catalogues are available online as supplementary material. The contents of the tables are given below, from columns (1) to (10):\\

\noindent
\textit{\textbf{Column (1)}}: Serial number assigned to each \H detection, based on which optical properties can be correlated with Tables \ref{tab:A963_optical_table} and \ref{tab:A2192_optical_table} for the two volumes respectively.

\noindent
\textit{\textbf{Column (2)}}: Galaxy ID based on the Right Ascension (J2000) and Declination (J2000) of the \H\s centres derived from our observations.

\noindent
\textit{\textbf{Columns (3) \& (4)}}: Right Ascension and Declination of the \H\s centres (J2000).

\noindent
\textit{\textbf{Column (5)}}: \H\s redshift $z_{\mathrm{HI}}$.

\noindent
\textit{\textbf{Column (6)}} : Luminosity distance $\mathrm{D_{lum}}$ in Mpc to the galaxy derived from co-moving distances based on the \H\s redshifts and the adopted cosmology. 

\noindent
\textit{\textbf{Column (7) \& (8)}}: \H\s linewidths $\mathrm{W^{obs}_{20, R4}}$ and $\mathrm{W^{obs}_{50, R4}}$ measured in km s$^{-1}$ at 20\% and 50\% of the peak \H\s flux, respectively, derived from the R4 cubes. These line widths have not been corrected for instrumental broadening.

\noindent
\textit{\textbf{Column (9)}}: Primary beam corrected integrated \H\s flux density S$\mathrm{_{int}}$ in mJy km s$^{-1}$ obtained from the R4 cubes.

\noindent
\textit{\textbf{Column (10)}}: The total \H\s mass in M$_{\odot}$.

\noindent
\textit{\textbf{Column (11)}}: Type of profile. As mentioned in Sect. \ref{gp}, all \H\s sources have been classified into three categories based on their global \H\s profiles: Double horned (Type 1), single Gaussian-like (Type 2) and asymmetric (Type 3).

\begin{table}
\begin{center}
\begin{tabular}{cccc}
\hline \hline
Date          & Used  & Seeing   & Weather  \\
& exposures & & conditions \\\hline
17 April 2007 & 20             & 1.0 - 1.2        & cirrus             \\
18 April 2007 & 12             & \textgreater 1.1 & clouds             \\
19 April 2007 & 25             & 1.2 - 1.4        & good               \\
04 April 2008 & 3              & 1.2 - 1.4        & clouds             \\
05 April 2008 & 0              & -                & high humidity      \\
06 April 2008 & 0              & -                & storm              \\
26 April 2009 & 0              & -                & high humidity      \\
27 April 2009 & 29             & \textgreater 1.4 & cirrus             \\
28  April 2009 & 31             & \textless 1.1    & good               \\ \hline
\end{tabular}
\end{center}

\caption{Exposures used in the final mosaic images along with their seeing conditions.}
\label{tab:optseeing}
\end{table}

\section{INT Wide field optical imaging}\label{wf}
Both fields have been imaged by the Sloan Digital Sky Survey \citep[SDSS DR7,][]{Abazajian09}. However, as the SDSS has limited surface brightness sensitivity and probes mostly the brightest galaxies on the red sequence at these redshifts, optical counterparts for some of the fainter \H detections were not available. Identification of these missing optical counterparts is important, however, for obtaining rough optical morphologies of the \H\s detected galaxies to probe the nature of the galaxies in our sample in relation to the BO effect. For this purpose, wide-field optical Harris images of the two \b\s fields were obtained with the Wide Field Camera (WFC) on the 2.54m Isaac Newton Telescope (INT) at the "Observatorio de la Roque Muchachos", La Palma. As a motivation for carrying out this study, Fig. \ref{fig:sdssvsint} shows the central region of A963 in both, the SDSS and the final INT composite images. From the figure, it is evident that the INT image is significantly deeper than SDSS, which is essential for subsequent studies.

The WFC consists of four CCDs with 2148 $\times$ 4128 pixels of 13.5 microns, corresponding to 0.33 arcseconds. The largest gap between the chips is 1098 microns, corresponding to $\sim$ 27 arcseconds on the sky. For both fields, the entire FWQM of the WSRT primary beam was imaged in the Harris B- and R- filters by mosaicking 30 individual exposures, distributed over 6 partially overlapping pointings. At each pointing position, 5 individual exposures were collected, slightly dithered with 35 arcsecond offsets around the centre of the pointing, in order to fill the gaps between the chips. The integration time for every individual exposure was 480s for the B-band and 360s for the R-band. The total integration time per field amounted to 14400 seconds or 4 hours for the B-band and 10800 seconds or 3 hours for the R-band imaging.

In total, 9 observing nights were allocated in April 2007, 2008 and 2009, of which 6 yielded usable scientific data. A summary of the number of exposures used in the making of the moisaic images and the atmospheric conditions during the observations is presented in Table \ref{tab:optseeing}. Bias frames and at least three sky flatfields in both filters were taken at the beginning and the end of each night. The dark current (considerably less than 1 e$^-$/hour) was neglected during the data processing. The nights were non-photometric and no standard star fields were observed.

\subsection{Data processing}
The imaging data from the INT WFC were processed with the {\it ccdred} and {\it mscred} packages of IRAF \citep{Tody86}.

\begin{figure}
\includegraphics[width=\linewidth]{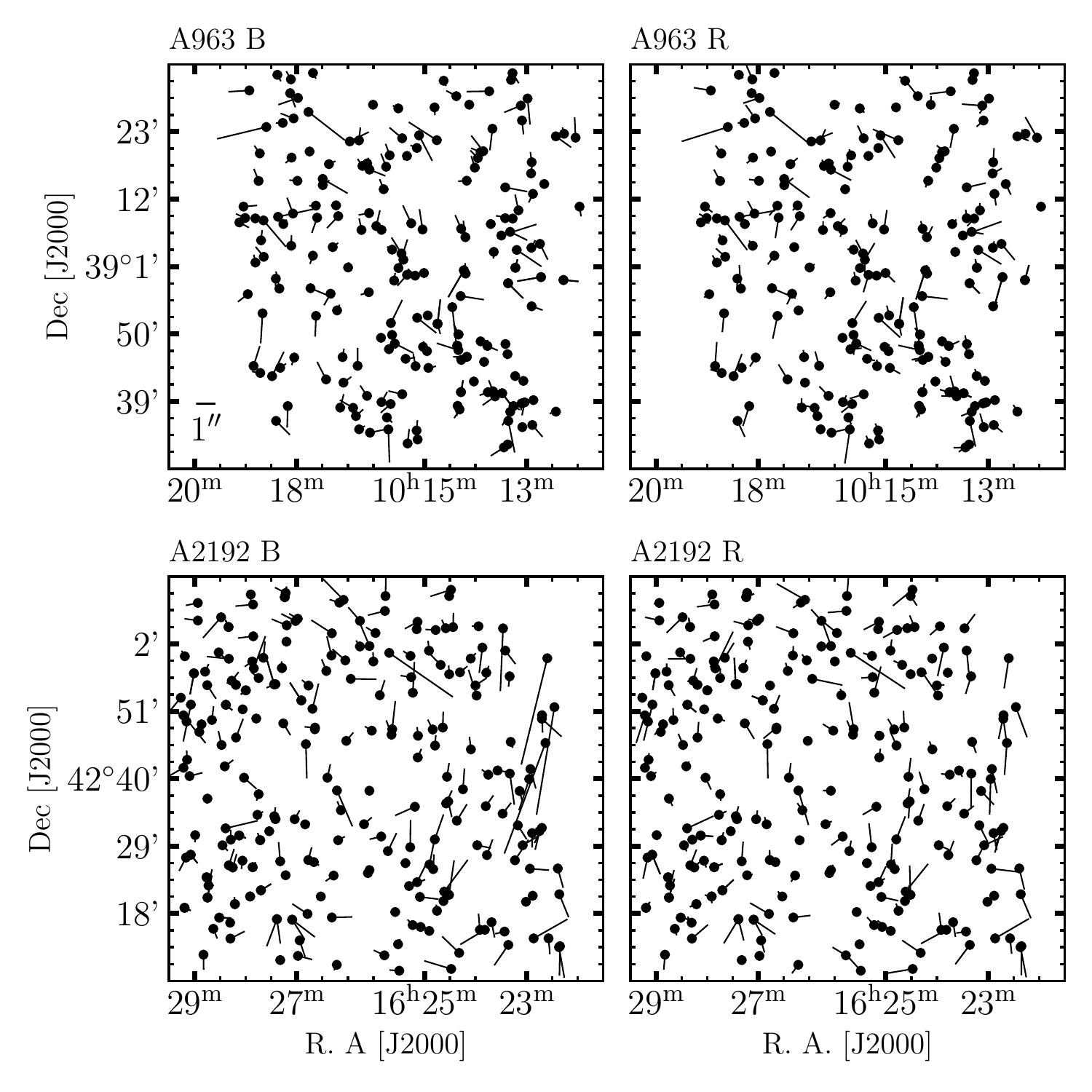}
\caption{For the 150 stars used for astrometric calibration, vectors showing the residual differences in the positions of the stars derived from our INT images and those from the USNO-A2.0 catalogue. These vectors have been enlarged by a factor of 250 for clarity. Top two panels: A963; bottom two panels: A2192. A size bar showing 1 arcsecond, also enlarged 250 times, is given in the bottom left corner of A963 B.}
\label{fig:astr}
\end{figure}

\begin{figure}
\includegraphics[width=\linewidth]{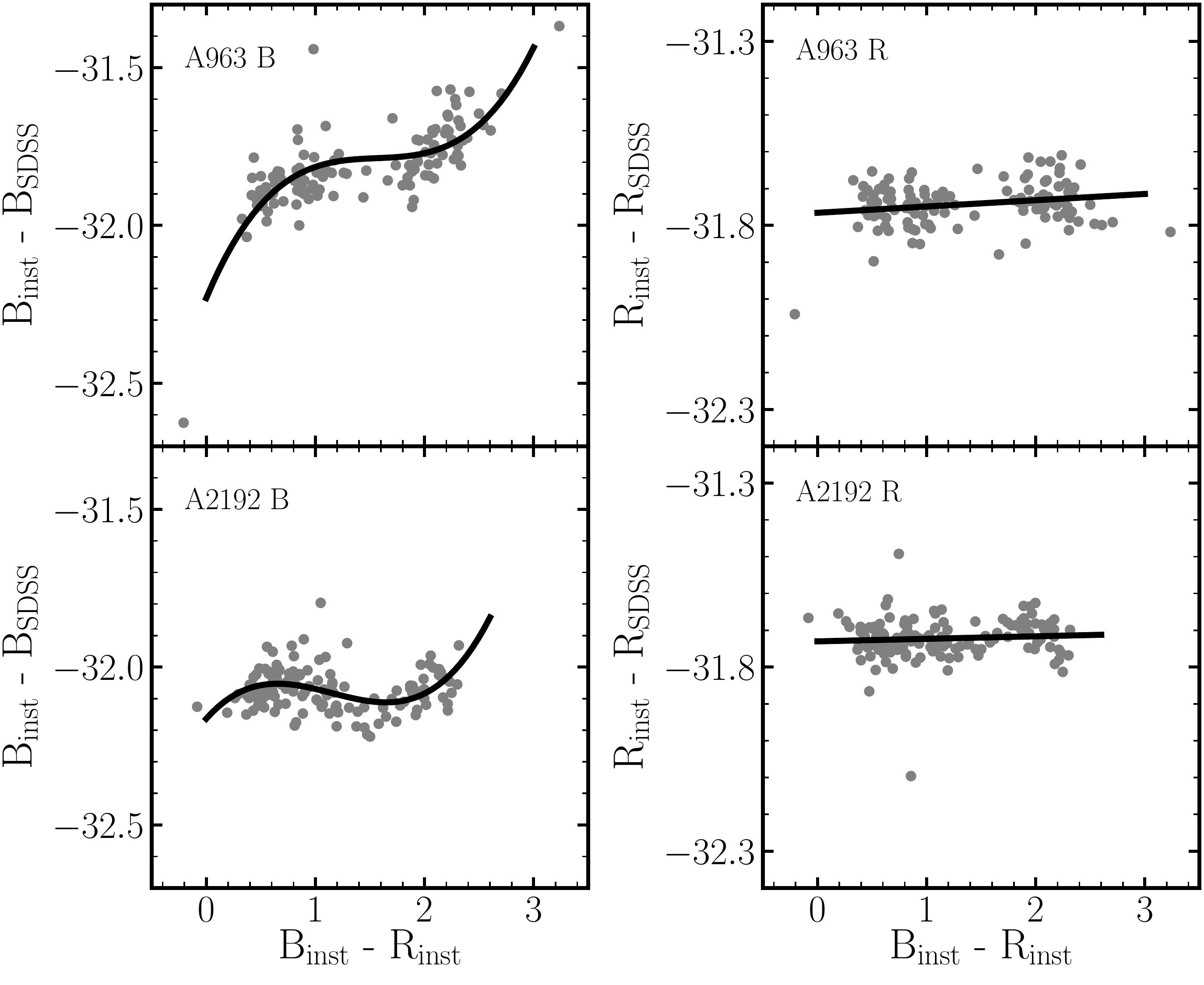}
\caption{Zero points derived from our photometric calibration for A963 (Top) and A2192 (Bottom). The grey points denote the zero points for both, the B - (left panel) and the R- bands (right panel). Horizontal axes show the instrumental colour derived from our images, whereas vertical axes indicate the photometric zero points, calculated as the difference between INT and SDSS magnitudes. Lines indicate polynomial fits made to the scatter points.}
\label{fig:zp}
\end{figure}

\begin{figure*}
\includegraphics[width=\textwidth]{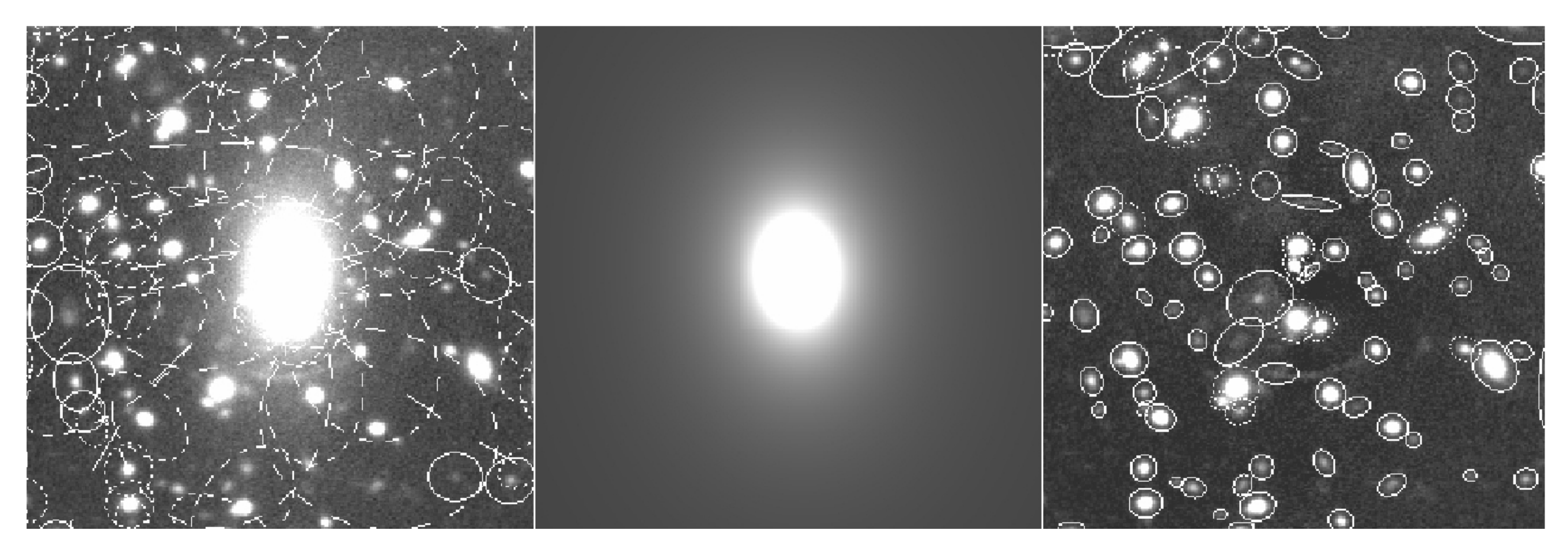}
\caption{Automatic apertures made with SExtractor before (left) subtraction of the model cD galaxy in A963 (middle), and  after (right) the subtraction of the model.}
\label{fig:se_cd}
\end{figure*}

\subsubsection{Bias removal, flat fielding and sky subtraction}
Each exposure yielded four "science frames" from the four CCD chips. The following procedure was carried out individually for every science frame from every exposure. The value of the overscan region was first subtracted, the frames were trimmed and a bad pixel mask was created. A median bias frame was then made per observing run and subtracted from all science frames and flatfields. A median of the normalized flats was made and used as a master flat field for each filter for each night. Every master flat image was then normalized to the highest median from all the four chips. All science frames were then divided by the corresponding flat field image.

For the sky subtraction, masks were made and grown to various sizes to censor the faintest galaxies and scattered light from bright stars. A single first order plane was then fit to the masked images and subtracted. The images were inspected interactively and any residual sky imperfection was then manually estimated and subtracted. A fringe correction was then made by subtracting the median of all the available R-band images from each individual R-band image. A preliminary coordinate system was attached to the images by aligning the catalogue stars taken from the NOAO USNO - A2.0 to the corresponding stars in the science frames. The astrometric solutions derived from this step were applied to all the individual science frames. 

All four science frames from a single exposure were combined into a single 4-CCD mosaic image, once the detector and filter induced artifacts were removed. An additional refinement of the coordinate system was carried out for each 4-CCD mosaic image to reduce any additional astrometric residuals, by re-aligning the NOAO USNO-A2.0 stars with those in the 4-CCD mosaic images.

Since the observations were carried out in non-photometric conditions and over a range of airmasses, the 30 4-CCD mosaics obtained through a particular filter for a particular field, 5 at each of the 6 pointing locations, had to be scaled to match the brightness of selected stars in the various overlap regions between the 30 4-CCD mosaics. First, the 5 dithered 4-CCD mosaics around a particular pointing location were scaled and combined into a single pointing mosaic to fill the gaps within each of the five individual 4-CCD images. Subsequently, the 6 pointing mosaics were scaled with respect to each other, again by using selected stars in their overlap regions. After this scaling the 6 pointing mosaics were combined into a single 1x1 deg$^2$ mosaic that covers the FWQM of the WSRT primary beam. This procedure resulted in four mosaiced images in total, one for each filter and each field. The photometric calibration and consistency across the mosaics will be discussed in Sect. \ref{phot}.

\subsubsection{Astrometric calibration}
The produced images were then further processed for an additional refinement of the astrometric calibration solutions. The images were first reprojected to match the projection centres of the \H\s maps produced by the WSRT for the two fields. About 150 stars taken from the USNO-A2.0 catalogue were identified in the INT images. The GIPSY task \textit{astrom} was used to measure any deviations and the corrections were subsequently applied using \textit{reproj}. Figure \ref{fig:astr} shows the amplitude and direction of the residuals in the astrometry for all the images at the positions of the 150 reference stars. These offsets are enlarged by a factor of 250 for clarity.  Note that we ignored the effect of proper motions of the stars in the astrometric solutions. We do not find systematic residuals in the plate scale, rotation angle or projection system.

\subsubsection{Photometric calibration}\label{phot}

For the purpose of photometric calibration, a catalogue of roughly 900 stars per field with 17.66 < r < 19.66 was extracted from the SDSS database. The lower magnitude limit was set to ensure that the stars are within the linear regime of the WFC CCD chips, while the upper limit ensured a sufficiently high SNR. Their  SDSS u, g, r and i magnitudes were then converted to the Johnson B- and Kron-Cousin R- magnitude filters based on the transformation equations (\ref{eq:B}, \ref{eq:R}, see \citet{Lupton05}). These filters match closely with the Harris B- and R- filters used for this study, and hence an additional conversion factor was not necessary.

\begin{equation}
\mathrm{B = u - 0.8116 \:  (u - g) + 0.1313} ; \: \: \sigma = 0.0095
\label{eq:B}
\end{equation}

\begin{equation}
\mathrm{R = r - 0.2936 \:  (r - i) - 0.1439} ; \: \: \sigma = 0.0072
\label{eq:R}
\end{equation}

\noindent
Out of the 900 stars in the catalogue, stars which were in relative isolation and with a uniform background in our INT images were identified and used for the aperture photometry. The remainder of the stars were masked out. Apertures were made around the stars out to a radius of 15 arcseconds. The outermost annuli were used for the estimation of the local sky background, which was subtracted from each of the other annuli. The annulus corresponding to the first dip in flux below the sky level was used as the outermost radius of the aperture. The measured fluxes were converted to instrumental INT B- and R- magnitudes derived from our images. Photometric zero points, calculated as the difference between the instrumental magnitudes and the converted SDSS magnitudes, were calculated for the two bands. Illustrated in Fig. \ref{fig:zp}, these zero points are plotted as a function of the instrumental colour of the INT data. For both fields and filters, a polynomial fit to the data showed a strong colour dependence in the B-band. The instrumental INT magnitudes of the galaxies were corrected following this calibration.

\subsection{Optical source finding}
We used SExtractor to identify and catalogue all the extended sources within the two fields. The program was run in dual mode, with the R-band image used as a detection image. The images were filtered with a Gaussian with a FWHM of 2.5 pixels. Sources were selected to have at least five contiguous pixels with values larger than 2$\sigma$ above the local background. The deblending parameter was set to zero since this provided the best results for cluster galaxies at z $\simeq$ 0.2. Those galaxies rarely showed any structures that would cause sextractor to fragment them. No other deblending setup was able to separate all the galaxies in the crowded central region of A963. The local background was estimated and subtracted using a mesh size of 128 pixels and a mean filtering box of 9 pixels. This adequately removed the extended halos around bright stars, which otherwise would have caused an erroneous estimation of the Kron radius and hence unreasonably big apertures for the photometry. Apertures close to the brightest stars affect sources out to a few arcminutes away and tended to be larger, thereby causing a systematic offset in the measured magnitudes. After subtracting the local background, these apertures became much smaller than before, but were nevertheless flagged in the final catalogue. In addition, the extended light from the cD galaxy in the centre of A963 was causing a similar problem for the aperture photometry of galaxies in the centre of the cluster. The cD galaxy was modeled with \textit{galfit} \citep{Peng02} and subtracted to allow for a better estimation of the Kron radii of the surrounding galaxies. Figure \ref{fig:se_cd} shows the automatic apertures in the centre of A963, before and after subtracting the cD galaxy. Not only were the automatic apertures more appropriate, but nearby sources were also better deblended. For all the objects in the catalogue, magnitudes were estimated in automatic apertures based on the default scaling parameters given by the SExtractor task \textit{mag\_auto}.

\begin{figure*}
\includegraphics[trim={0cm 3.2cm 0cm 3cm},clip,width=0.95\textwidth]{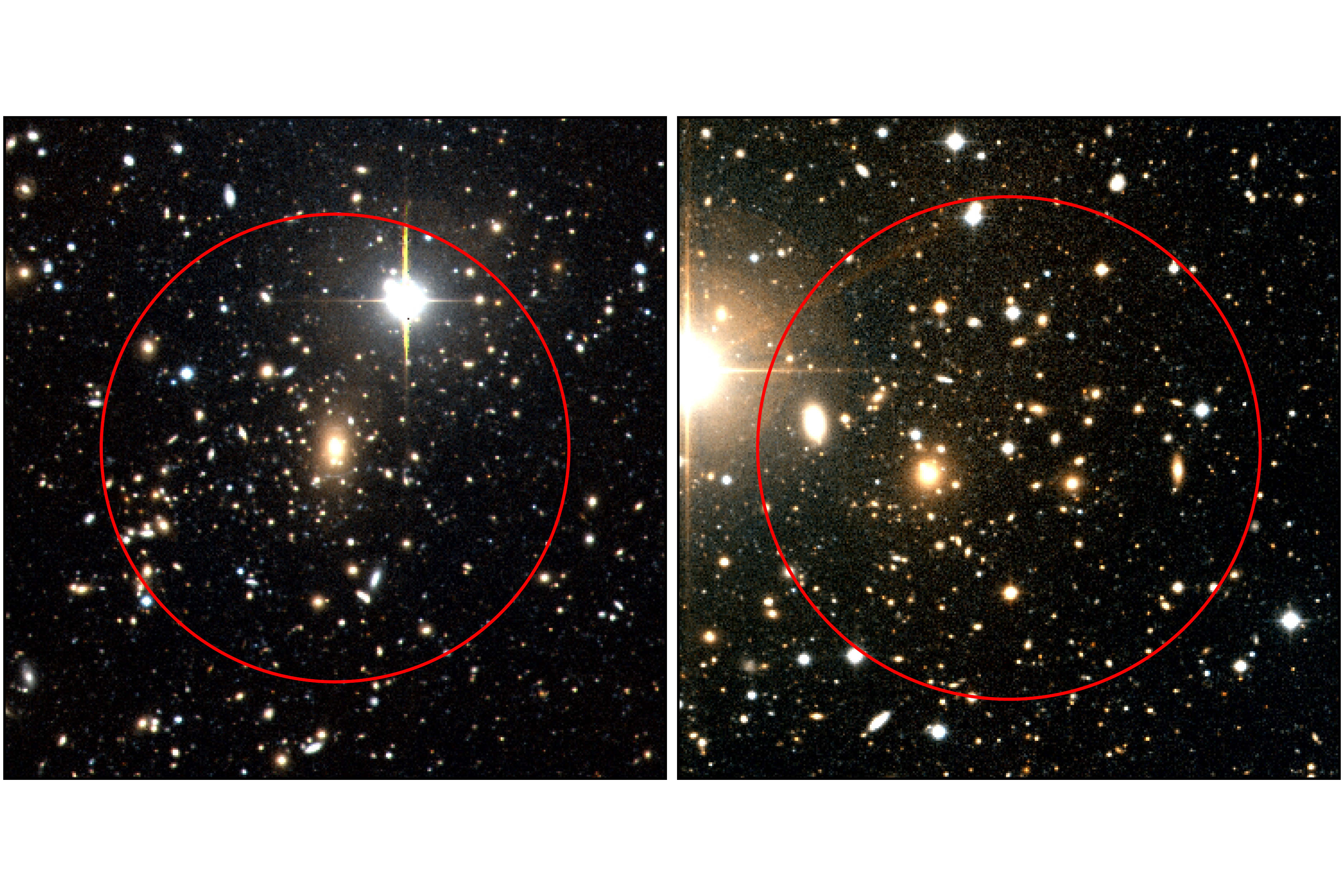}
\caption{Optical images constructed from the B- and R-band images of the two clusters A963 (left) and A2192 (right) at z $\simeq$ 0.206 and z $\simeq$ 0.188 respectively, taken with the WFC of the INT. The two images show the central 7$\times$7 arcmin$^2$ of the entire 1 degree$^2$ field surveyed by the INT. The red circles indicate the central 1 Mpc region in diameter centred on the two clusters at their respective redshifts.}
\label{fig:opticalimg}
\end{figure*}

The final B- and R- band magnitudes are provided in the optical source catalogues (Tables \ref{tab:A963_optical_table} and \ref{tab:A2192_optical_table}). Many low surface brightness sources, as well as arcs around the central cD galaxy can be seen in the INT image, which are absent in the SDSS image of the same field. Figure \ref{fig:opticalimg} shows the central 7 $\times$ 7 arcmin$^{2}$ of the final images of the two clusters as constructed from our INT data. 

\subsection{The optical catalogue}\label{optcat}
Tables \ref{tab:A963_optical_table} and \ref{tab:A2192_optical_table} list the derived optical properties for all confirmed \H\s detections in the two volumes with available INT or SDSS photometry. These optical counterparts are identified in either SDSS or our INT observations, or both. The contents of the tables are summarized as follows:\\ \\
\noindent
\textit{\textbf{Column (1)}}: Serial number corresponding to Tables \ref{tab:A963_HI_table} and \ref{tab:A2192_HI_table}.

\noindent
\textit{\textbf{Column (2)}}: SDSS Galaxy ID based on the Right Ascension (J2000) and Declination (J2000) of the optical centre. For those sources which did not have an SDSS counterpart, INT IDs have been provided.

\noindent
\textit{\textbf{Column (3) \& (4)}}: Right Ascension and Declination (J2000) of the optical centres derived from our INT observations.

\noindent
\textit{\textbf{Column (5)}}:  Position angle of the detected galaxies.

\noindent
\textit{\textbf{Column (6)}}: Optical redshift $z$ from the literature for those objects which have optical spectroscopy.

\noindent
\textit{\textbf{Column (7)}}: B-magnitudes for all sources calculated after INT data reduction and source extraction. These values have not been extinction or k-corrected.

\noindent
\textit{\textbf{Column (8)}}: R-magnitudes for all sources calculated after INT data reduction and source extraction. These values have not been extinction or k-corrected.

\noindent
\textit{\textbf{Column (9) \& (10)}}: Observed FUV and NUV magnitudes extracted from GALEX (see Sect. \ref{uv}).

\begin{figure*}
\includegraphics[width=0.9\linewidth]{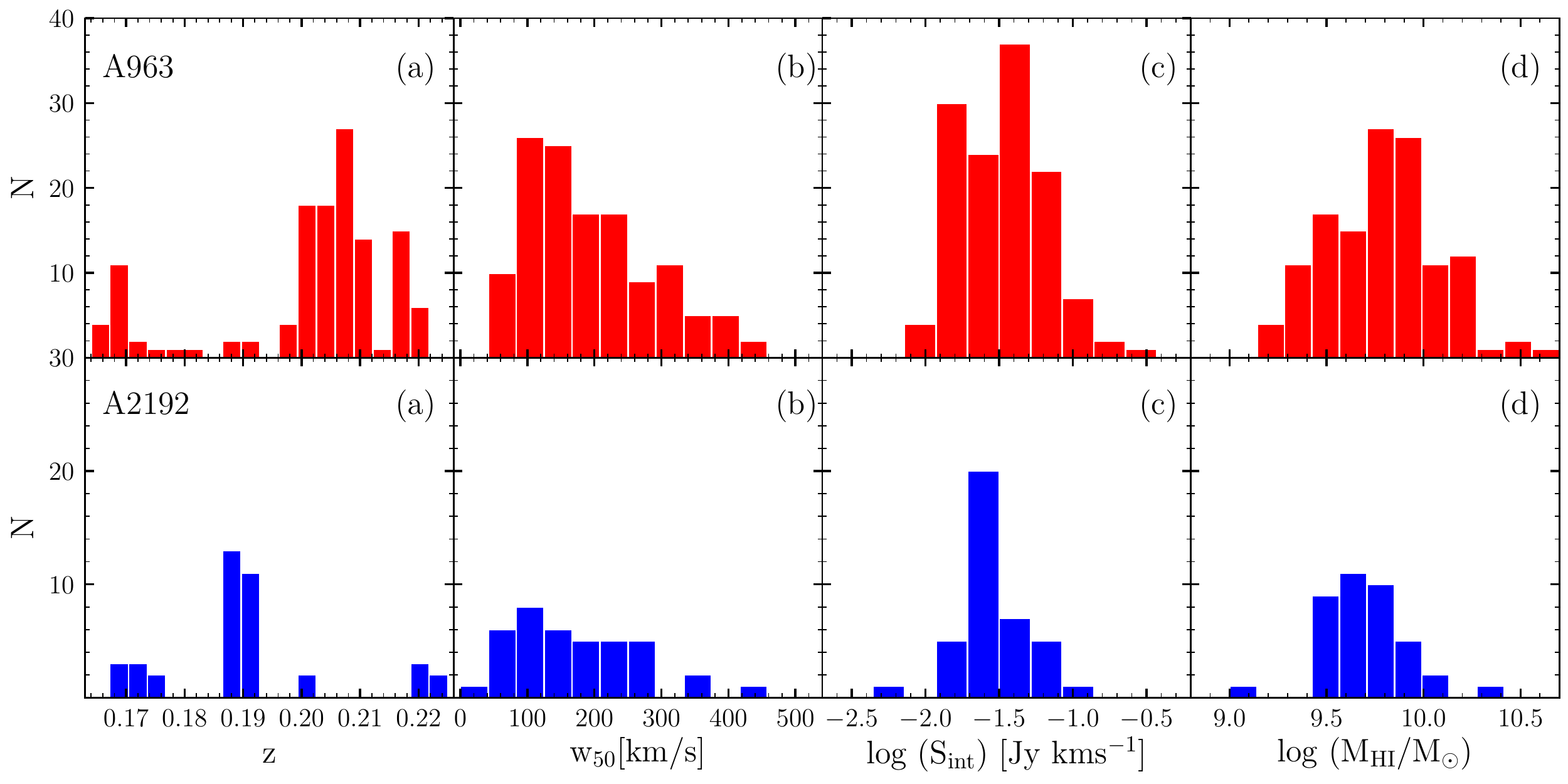}
\caption{Histograms illustrating various global \H\s properties of the detected galaxies derived from the R4 cubes shown as the red and blue histograms for A963 (top panel) and A2192 (bottom panel) respectively. From left to right: Redshift (a), \H line width (w$_{50}$) (b), integrated flux (c), \H mass (d).}
\label{fig:hists}
\end{figure*}

\section{GALEX - UV imaging}\label{uv}
While information on the \H content of blue galaxies is useful in the context of the BO effect, it is necessary to link the gas content with the recent SF History (SFH) of the galaxies that are responsible for the BO effect. To properly assess the impact of environmental effects on the SFH and evolution of galaxies it would be very helpful to resolve the recent SFH on time scales that are relevant to the environment dependent astrophysical processes that act upon the galaxies, where some of these processes such as ram pressure stripping can act on rather
short time scales. While the average SF activity over the past few 100 Myrs can be estimated in the \H\s detected blue galaxies using our optical photometry and radio continuum fluxes, the currently ongoing SF activity can only be accurately measured with the help of UV photometry as the UV luminosity of a galaxy is dominated by the short-lived but bright O- and B-type stars.  Therefore, we obtained deep UV imaging photometry of the two BUDHIES fields in order to help us disentangle the currently ongoing SF activity from the average SFR over the past few 100 Myrs in the \H detected blue galaxies. 

The Galaxy Evolution Explorer (GALEX) satellite, launched on 28 April 2003 and decommissioned on 28 June 2013, was used to simultaneously obtain deep near-UV (NUV, 1771-2831 \AA) and far-UV (FUV, 1344-1786 \AA) images for both fields separately.  GALEX provides a circular field-of-view with 1.24 and 1.28 degree diameters for the NUV and FUV respectively, which is a good match to the FWQM of the WSRT primary beam. The angular resolution of the GALEX images is 5.3 arcsec in the NUV and 4.3 arcsec in the FUV.  The pointing centres, observation dates, number of orbits and exposure times for the two fields are given in Table \ref{uvtab}.  The achieved limiting magnitude for point sources in both the NUV and FUV was m$_{AB}$ $\sim$ 25.5 mag.  The UV images and photometry reported in this paper are produced with the GALEX Pipeline, which uses SExtractor to measure magnitudes by applying a variable extraction area (aperture) depending on the object size.  A detailed explanation of the data products from the GALEX Data Analysis Pipeline can be found in \citet{Morrissey07}. 

Around 400 UV sources in A963 and 300 UV sources in A2192 were detected. This is strikingly more than the number of \H detections in the two fields but, obviously, the \H detections are restricted to the redshift range surveyed by the WSRT while GALEX also detected foreground and background galaxies.  The correspondence between the GALEX and radio continuum sources will be investigated in a forthcoming paper. Interestingly, we noted that many of our \H\s detections are also clearly detected in the GALEX images.  This allowed us to take advantage of the UV images for counterpart identification as discussed in Sect. \ref{sf}. The measured FUV and NUV magnitudes for most of our \H detections are presented in Tables \ref{tab:A963_optical_table} and \ref{tab:A2192_optical_table}.  The NUV images corresponding to the \H\s detections are included in the atlas (see Sect.  \ref{atlas}). Apart from two \H sources in A963 which lie just outside the GALEX field-of-view, all \H sources have optical and UV counterparts, most of them with confirmed optical redshifts.  Nearly all \H detections have a NUV counterpart and only a few are missing a FUV detection.  Occasionally, only an FUV counterpart for an \H detection could be identified.

\begin{table}
\centering
\begin{tabular}{ccc}

\hline \hline
 & A963 & A2192 \\ \hline
RA (J2000)  &10h 17m 09.6s  & 16h 26m 37s \\
Dec (J2000)  & 39d 01m 00s & 42d 40m 20s  \\
Date & 15 Feb 2008& 26 May 2008 \\
n$_{orb}$ &24 & 29 \\
T$_{exp}$ (s) & 29040 & 25815 (NUV)\\
& & 25796 (FUV)\\\hline
\end{tabular}
\caption{Table with details on the pointing centres, dates of observations, number of orbits per field and the exposure times for the GALEX UV observations of the two fields.}
\label{uvtab}
\end{table}

\begin{figure*}
 \includegraphics[trim={2cm 0.5cm 0.4cm 1cm},clip,width=\textwidth]{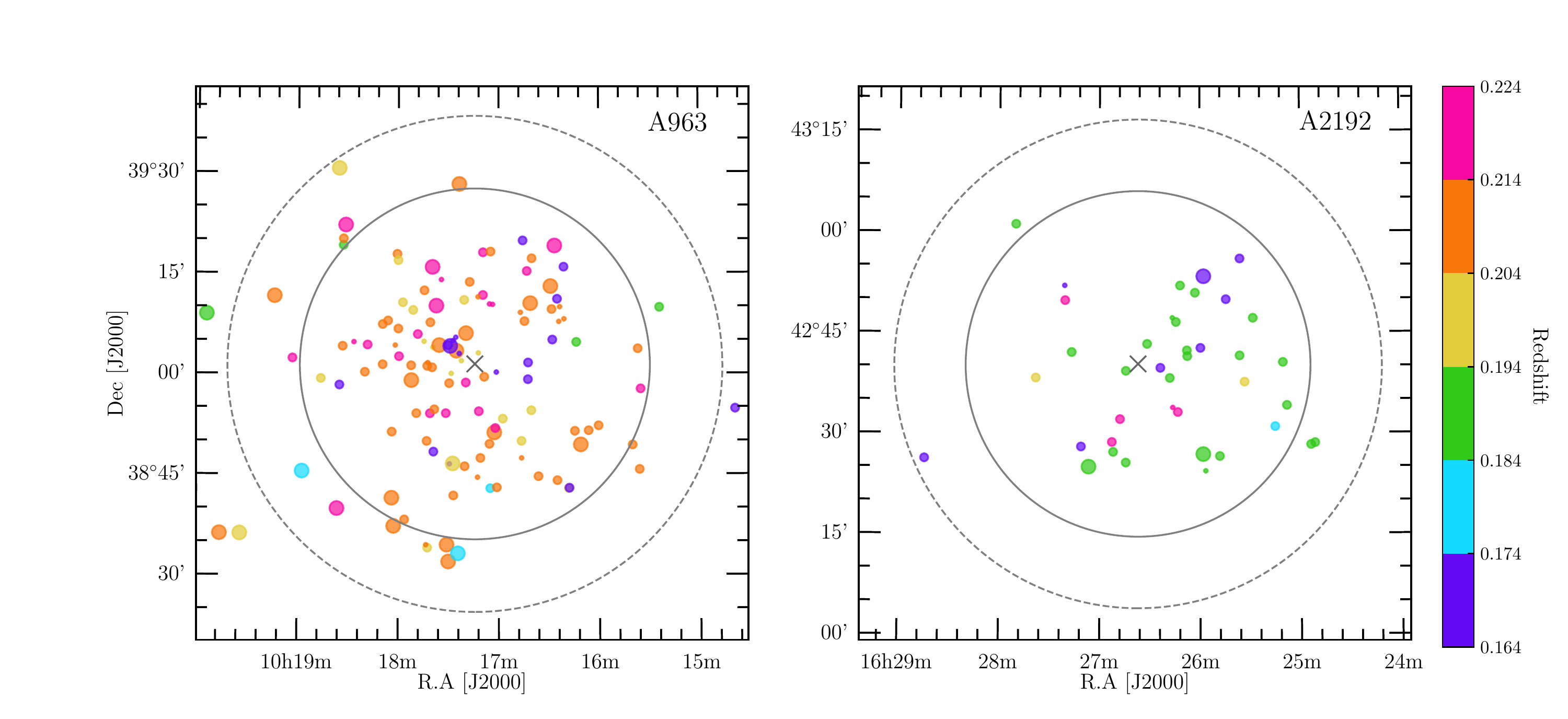}
 \caption{Distributions on the sky of the detected \H\s sources in the two volumes; Left: A963, Right: A2192. The distribution is colour coded based on redshift, as indicated by the colour bar, whereas the three marker sizes indicate three mass ranges of the galaxies: massive (\lgm > 10), intermediate (9.5 < \lgm < 10) and low (9 < \lgm < 9.5). The pointing centres for the two fields are indicated by the grey crosses. The FWQM and FWHM of the primary beam at cluster redshifts are indicated by the outer (dashed) and inner circles respectively.} 
\label{fig:spatdist}
\end{figure*}
\section{Results}\label{results}
In this section, we bring forth a discussion on some galaxy properties derived from a preliminary analysis of the various data sets.

\subsection{Distribution of HI properties}\label{prop}
To study the distribution of the global \H\s properties of the detected \H\s sources, we present plots of the various observed and derived \H\s parameters in this section.

Figure \ref{fig:hists} shows the distribution of \H\s redshifts, intrinsic line widths, integrated \H\s fluxes and derived \H\s masses of the galaxies. The large scale structure along the line-of-sight is clearly evident in both volumes in  Fig. \ref{fig:hists} (a) and has been described briefly in Sect. \ref{targets}. For further details, we refer to \citet{Yara1_13}.

Figure \ref{fig:hists} (b) shows histograms of the intrinsic W$_{50}$ line widths of galaxies which range from 61 to 451 km s$^{-1}$ for A963 and 26 to 453 km s$^{-1}$ for A2192. The average line widths for galaxies in A963 and A2192 are 193 and 172 km s$^{-1}$ respectively. The different distributions of the line widths for the two volumes could be due to differences in sensitivity, large scale structure, galaxy populations and sample sizes, A963 being statistically more significant than A2192.

Figures \ref{fig:hists} (c) and (d) show the histograms of the integrated \H\s fluxes (Jy km s$^{-1}$) and the derived \H\s masses (M$_\odot$) for the galaxies in the two volumes.  The integrated \H fluxes span a range between 7.5 to 229 mJy km s$^{-1}$ for A963 and 5 to 124 mJy km s$^{-1}$ for A2192. The average integrated flux values are 42 and 34 mJy km s$^{-1}$ for A963 and A2192 respectively.

The \H\s masses of the galaxies in the survey lie within a mass range \lgm  = 9.2 to 10.64. The average \lgm\  are 9.8 and 9.7 for A963 and A2192 respectively. The distribution appears to peak around the local $\mathrm{M_{HI}^{*}}$ \citep{Jones18} for A963 but is slightly shifted towards lower masses for A2192.  The significance and reason for this shift will be investigated with the help of \H\s mass-to-light ratios of the galaxies in an upcoming paper since it is beyond the scope of this publication. There also seems to exist a tail towards higher \H masses, which could be associated with galaxy mergers or interactions, or compact galaxy groups. Due to the large synthesized beam of the WSRT at these redshifts, such systems may not be resolved and hence a more detailed inspection of their optical counterparts and global profiles needs to be carried out.

Figure \ref{fig:spatdist} illustrates the distribution on the sky of all the \H sources in the two fields. They are colour coded based on their redshift and sized based on their \H mass. Note that most of the lowest \H\s masses are detected only near the centre of the two volumes within the FWHM of the primary beam at the redshifts of the clusters, while a few gas-rich systems are detected beyond the FWQM of the primary beam. This is due to the primary beam attenuation of the telescope, causing sensitivity to drop as a function of distance from the centre. The spatial distribution of A963 is relatively symmetric, whereas A2192 shows clear asymmetry, with a majority of the \H\s detections to the west of the pointing centre. From their colour-coding and Fig. \ref{fig:hists}(a), it is evident that most of the detected galaxies have redshifts similar to the clusters within the redshift range 0.204 < z < 0.211 for A963 and 0.188  < z < 0.192 for A2192.

\begin{figure*}
 \includegraphics[width=0.49\linewidth]{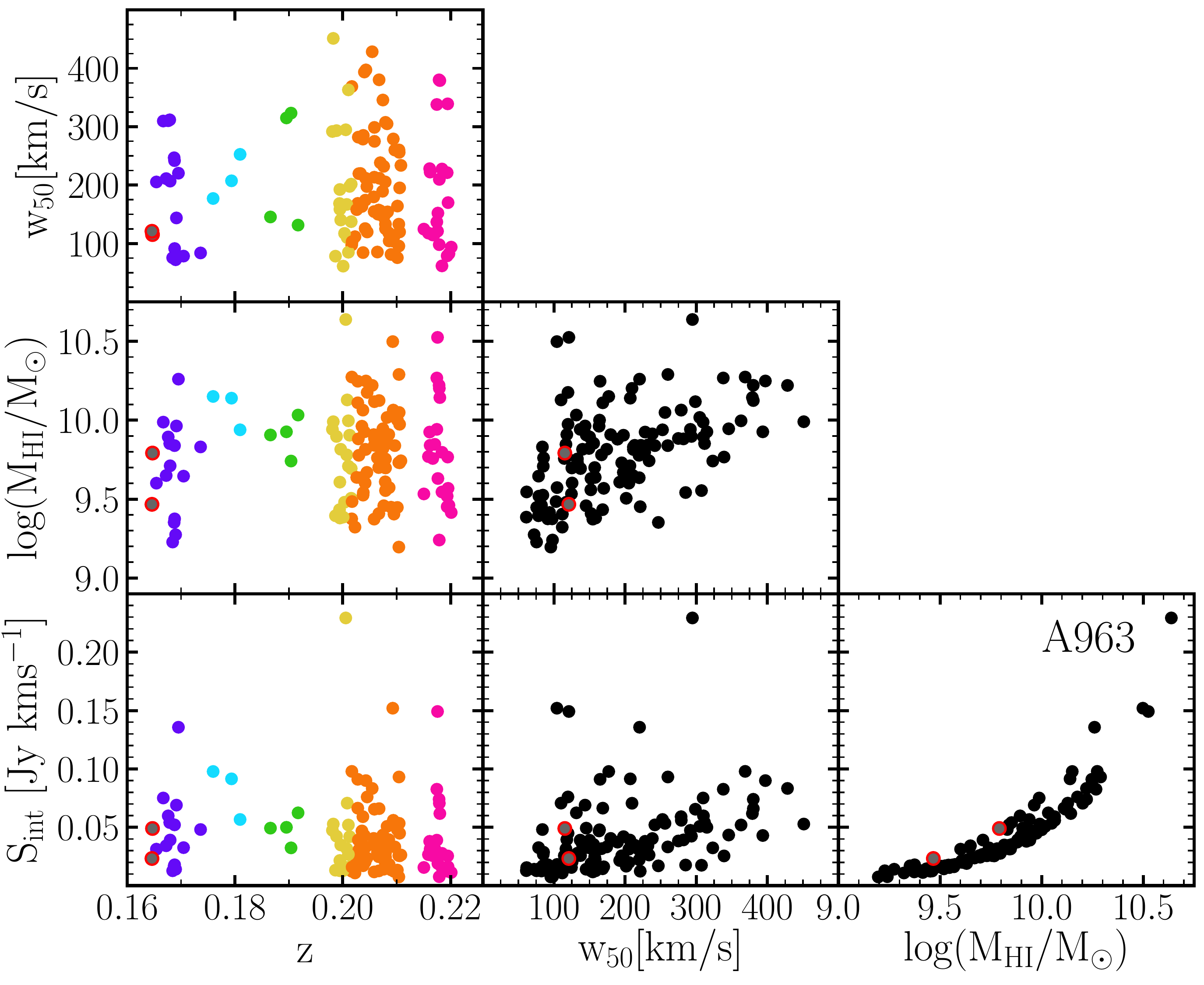}
 \includegraphics[width=0.49\linewidth]{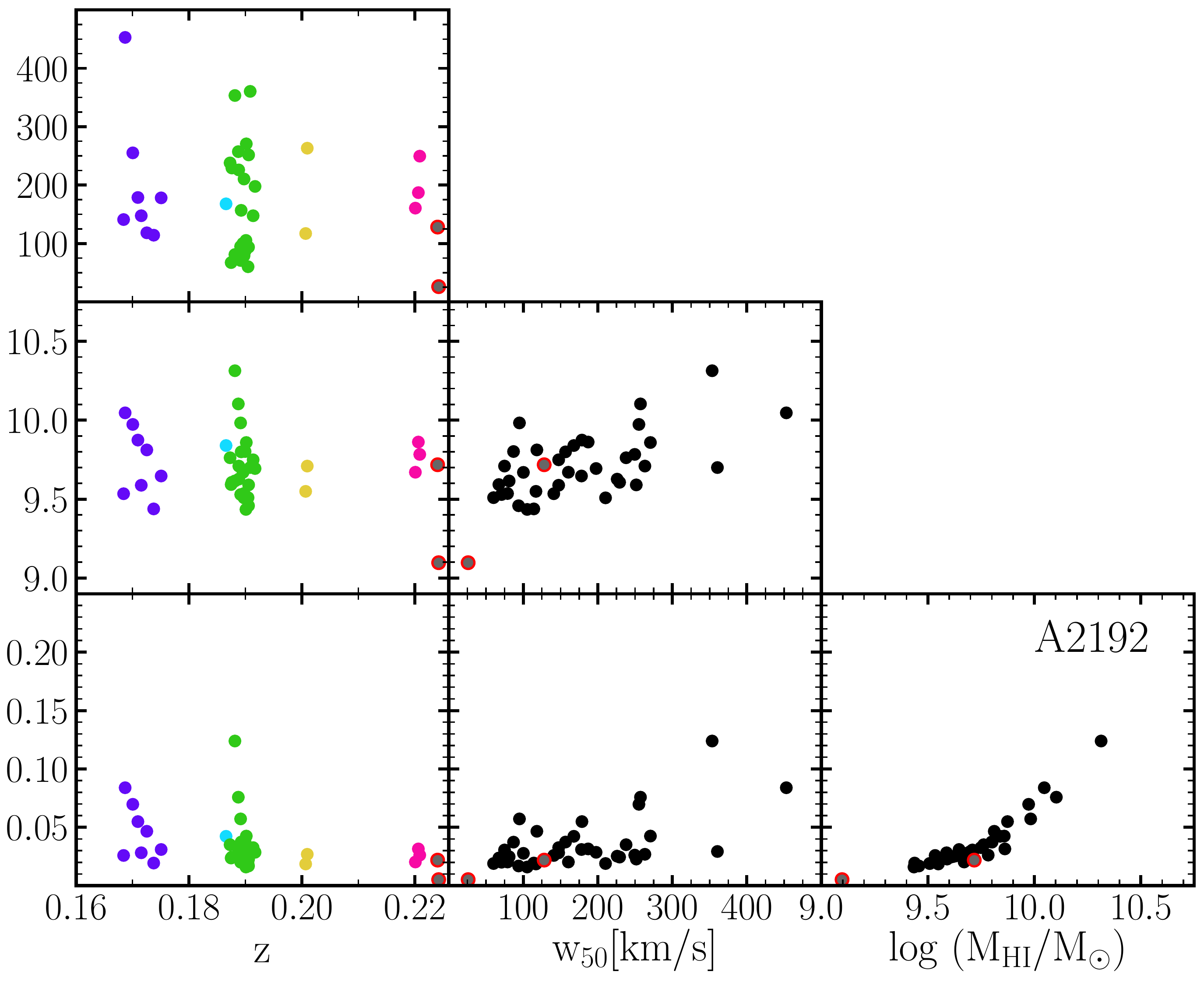}
 \caption{Multivariate distribution of the observed and derived \H properties, namely redshift (z), line width (W$_{50}$), \H mass (\lgm) and integrated flux \H\s (S$_{\mathrm{int}}$) for the two volumes. Left: A963; Right: A2192. The colours in the first (vertical) panel in both figures are based on the same colour distribution as in Fig. \ref{fig:spatdist}. The grey sources encircled in red are to be treated as lower limits, since they are very close to the band edges or bright continuum sources.}
 \label{fig:bivar}
\end{figure*}

 Figure \ref{fig:bivar} shows multivariate distributions of the derived \H parameters for the two volumes. The left column with three panels shows the distribution of the intrinsic \H\s line widths, inferred \H\s masses and integrated \H\s fluxes as a function of redshift. The colours follow the same colour bar as Fig. \ref{fig:spatdist}. There exists a large void in the foreground of A963 and a smaller void in the background. In the case of A2192, there appear to be smaller overdensities  separated by voids from the main cluster at z $\simeq$ 0.188. The large scale structure in the two volumes has been studied in detail by \citet{Yara1_13}. The middle column with two panels shows the distribution of integrated \H fluxes and \H masses as a function of their line widths (W$_{50}$), while the last column shows the integrated fluxes as a function of \H mass. Despite the modest \H mass and redshift ranges, trends in the various distributions are clearly visible. From Fig. \ref{fig:bivar}, a clear trend is observed in W$_{50}$ as a function of mass. It is observed that only the lower-mass galaxies with narrower line widths were detected, while galaxies with higher masses were found over a larger range of line widths. Since the \H\s line widths of galaxies depend on factors such as their inclinations and rotational velocities, the undetected low-mass galaxies with higher line widths in the bottom-right of this panel could be fast-rotating early-type gas-poor galaxies. As mentioned in the previous section, the detection limit of the survey was 2 $\times$ 10$^9$ \ms at the distances of the two clusters over a line width of 150 km s$^{-1}$. However, we still notice galaxies below the mass limit being detected at those redshifts and beyond. This is caused by line widths narrower than 150 \kms, which brings the detection limit down to lower masses. In the case of A2192, the galaxy with the lowest \H\s mass lies at the edge of the bandpass, and has a corresponding line width of about 15 km s$^{-1}$. Such galaxies were cut at the edge of the bandpass and only show a small section of their line profiles and hence low \H\s masses, and are to be treated as lower limits. These could also be some of the possible causes for the difference observed in the peaks of the histograms seen in Fig. \ref{fig:hists} (d).

No \H\s was detected in galaxies within a projected distance of 1 Mpc of the cluster A963, whereas for A2192 we found one \H\s source. As mentioned in Sect. \ref{intro}, A963 being a BO cluster, it is essential to know the nature of the blue galaxies using information on the gas content from this survey. Using the pilot sample, \citet{Verheijen07} stacked the \H spectra of the blue galaxies in the core of both clusters and compared them with the stacked spectra of the blue galaxies within and outside the central 1 Mpc region. They concluded that the central blue galaxies are significantly more \H\s deficient than those outside this region. A further analysis of the stacked spectra of galaxies with available optical redshifts inside and outside R$_{200}$ of A963 was carried out by \citet{Yara3_16} using the entire dataset. They found that the \H\s content of galaxies  in the cluster core, inside R$_{200}$, was reduced by half compared to those in the outskirts. They also concluded from stacking exercises in phase-space that the blue galaxies in the centre of A963, which are possibly infalling from the field, have been stripped of a large fraction of their \H\s during their transition into the stripping zone, i.e. their first passage through the ICM. Most of these blue galaxies were not detected in \H\s individually as any remaining \H\s in these galaxies is below our detection threshold.
In an upcoming paper, we will revisit the results for the \H\s stacking and relative gas content in terms of a comparative analysis of A963 and A2192, thereby building on the works of \citet{Verheijen07} and \citet{Yara3_16} on an \H\s perspective of the BO effect, using the full data set.

\begin{figure}
\includegraphics[width=\linewidth]{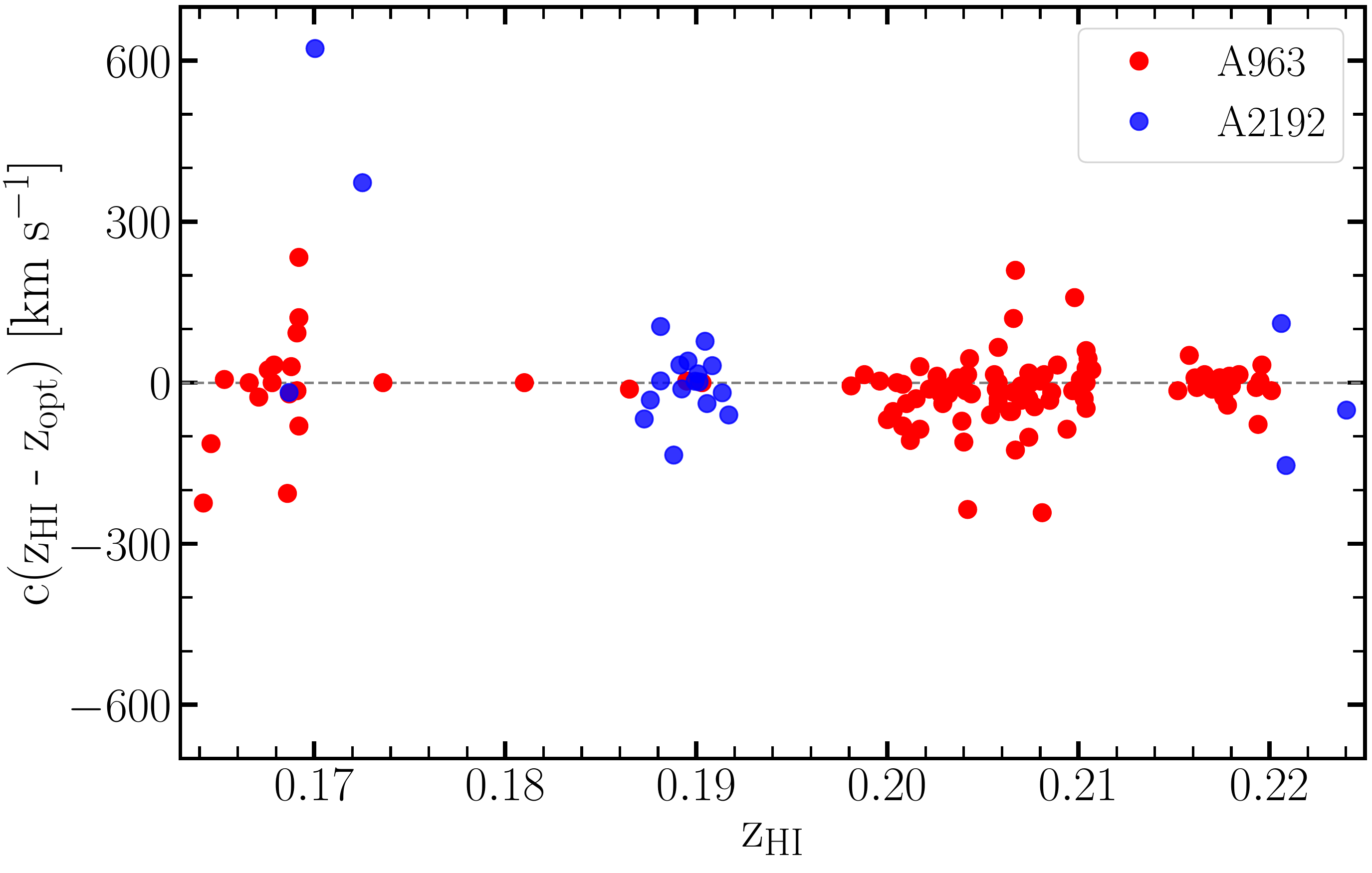}
\caption{For those sources with literature spectroscopic redshifts \citep[SDSS,][]{HL84, Yara1_13}, a plot showing a comparison between literature redshifts as a function of \H redshifts obtained from this survey.}
\label{fig:zcomp}
\end{figure}

\begin{figure*}
\begin{center}
\includegraphics[width=0.70\linewidth]{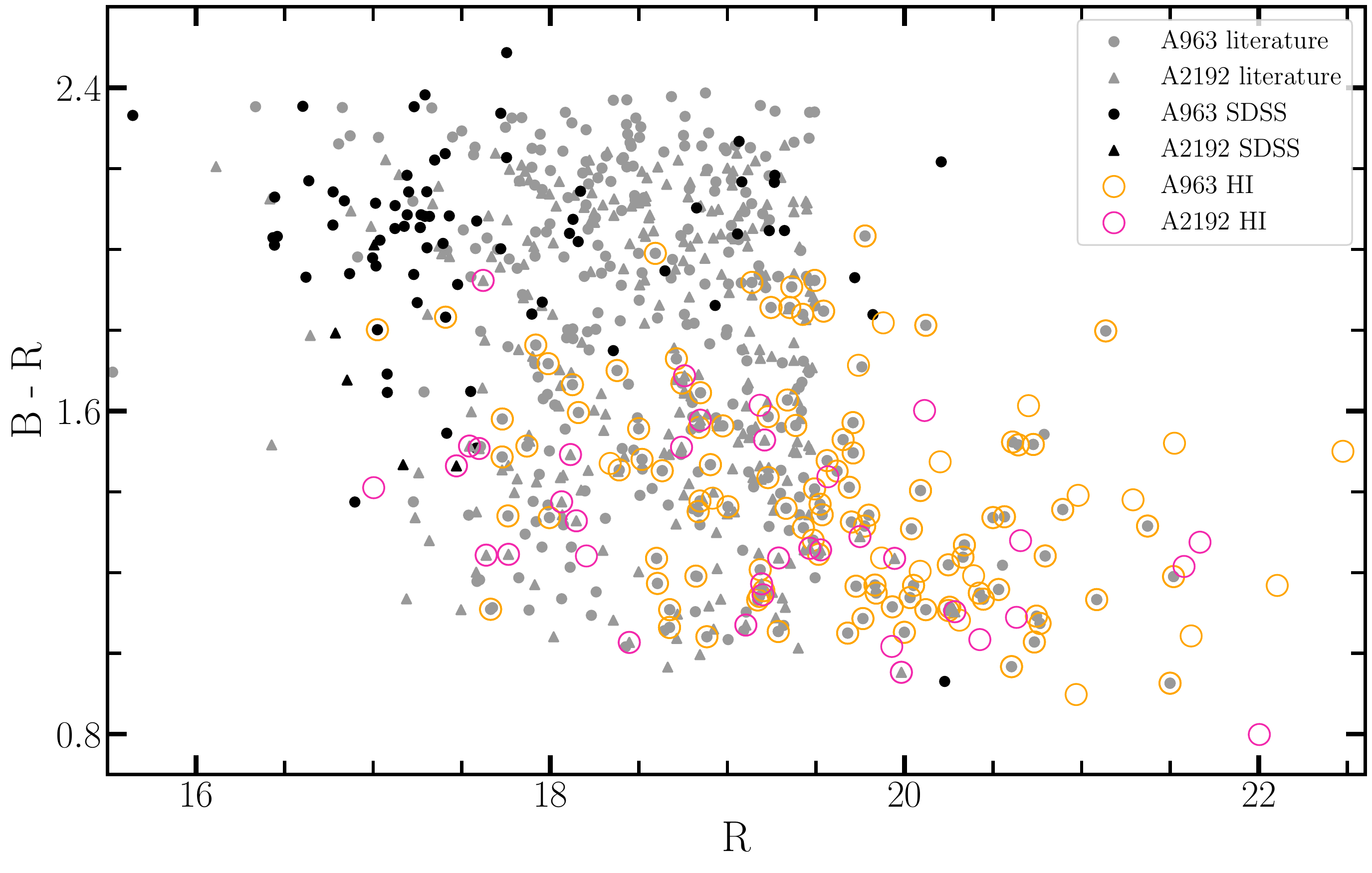}
\caption{The combined colour-magnitude diagram of all galaxies lying in the redshift range 0.164 < z <0.224. Galaxies with optical redshifts are indicated in grey, while those from SDSS are marked in black. These galaxies are mostly located within the "red sequence". \H\s detections are indicated as open circles. Most of the \H\s detections are located within the "blue cloud".}
\label{fig:cmd}
\end{center}
\end{figure*}

\begin{figure}
\includegraphics[trim={3cm 15.4cm 6cm 6cm},clip,width=\linewidth]{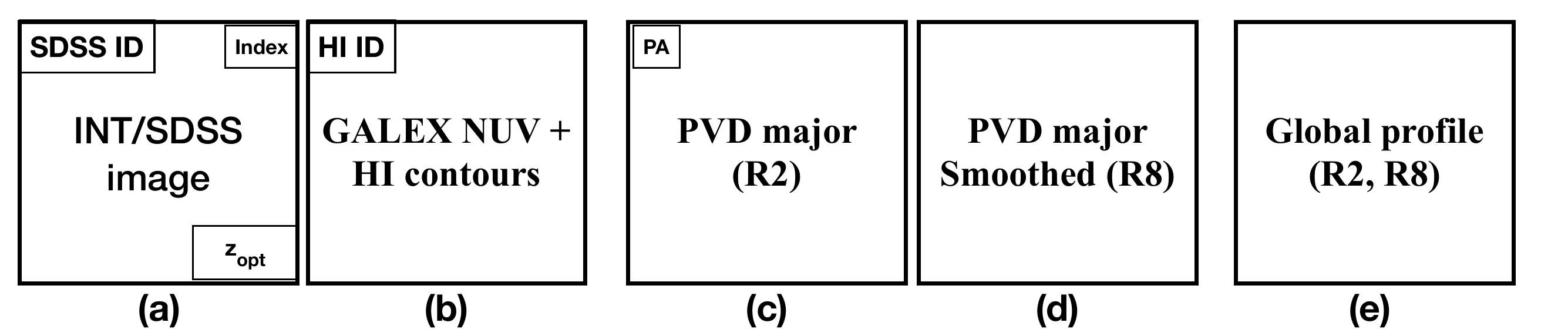}
\caption{Schematics of the arrangement of the derived \H\s parameters given in the atlas.}
\label{fig:lyt}
\end{figure}
\subsection{Optical properties}
Figure \ref{fig:zcomp} compares the \H redshifts with optical spectroscopic redshifts available in the literature \citep[SDSS,][]{HL84, Yara1_13}. The plot shows a tight correlation between the two redshifts with a small scatter of 70 \kms and 168 \kms for A963 and A2192 respectively. The scatter for both volumes combined is 94 \kms. This scatter becomes particularly significant for stacking exercises as well as for TFr studies, as it indicates the extent of broadening that can be expected in the width of the stacked profiles due to errors in the optical redshifts.

Figure \ref{fig:cmd} illustrates the B - R versus R colour-magnitude diagram (CMD) of all galaxies in the two volumes that have optical or \H\s redshifts. These colours and magnitudes have not been corrected for Galactic extinction or reddening.  A similar CMD was presented by \citet{Yara1_13} for the two clusters separately, while \citet{Yara3_16} presented the NUV - R versus R CMD of the cluster A963. From Fig. \ref{fig:cmd}, it is evident that most of the galaxies with SDSS redshifts lie at the bright tip of the red sequence, while most of the \H\s detected galaxies are located in the 'blue cloud'. However, we find that some \H\s detected galaxies lie close to the red sequence, in the so-called 'green valley'. The brightest \H\s detected galaxies are found to be as bright as the most luminous red galaxies, and could possibly be merging or post-starburst systems. A further study of the optical morphologies of all these galaxies is required in order to study these bright blue gas rich galaxies in greater detail.

\subsection{The atlas}\label{atlas}
\begin{landscape}
\begin{figure}
    \includegraphics[scale=1.2, trim={3cm 0cm 2cm 0cm},clip]{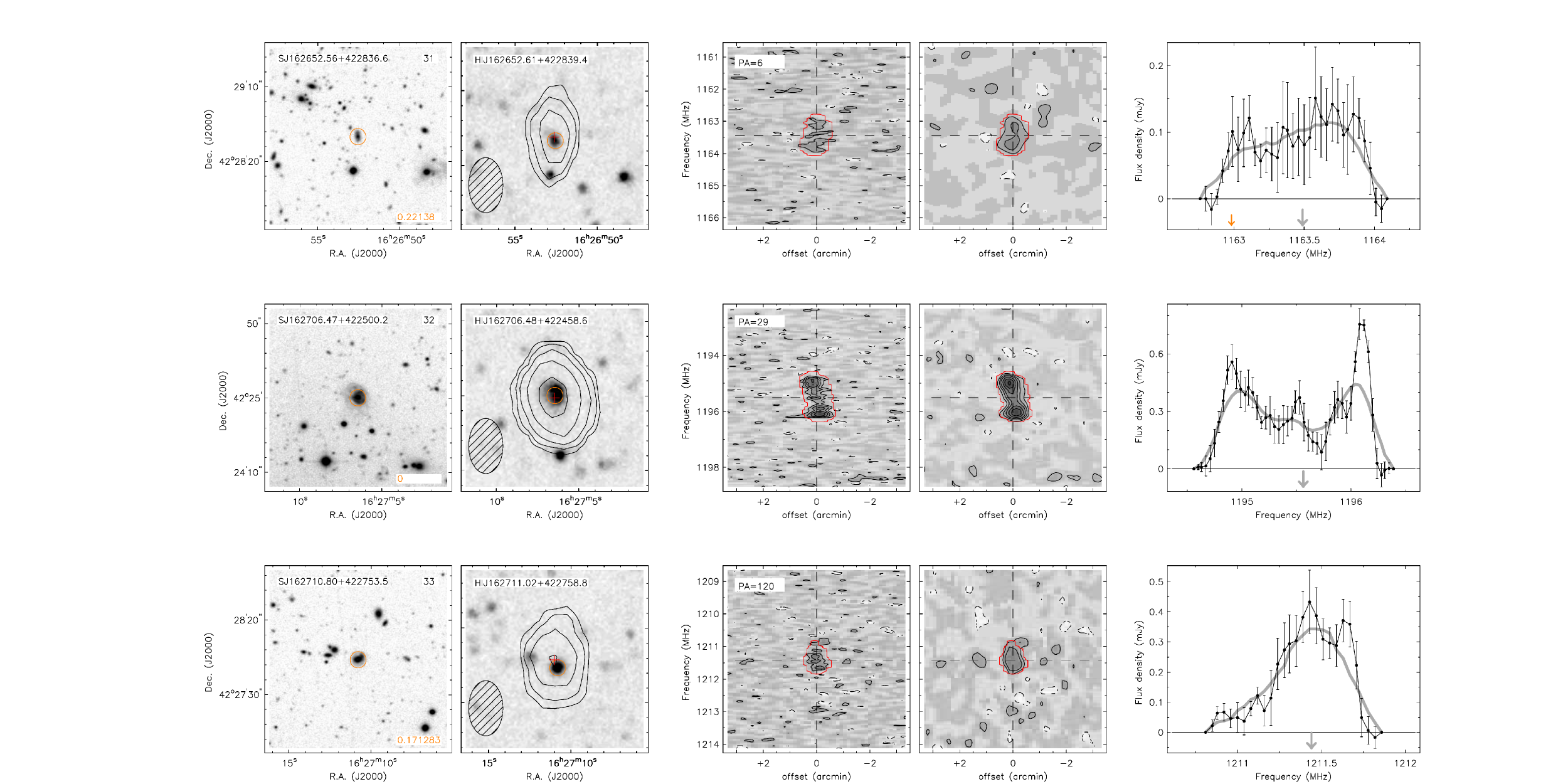}
    \caption{A sample atlas page showing the data products for three of the \H\s detections in A963. The full atlas is available online.}
    \label{fig:atlasA963}

\end{figure}

\end{landscape}

The \H\s atlas illustrates the most important \H\s properties described in Sect. \ref{gloprop} for each individual detected galaxy along with their INT R-band image (see Sect. \ref{wf}),  and ultra-violet GALEX image (see Sect. \ref{uv} for details).  Figure \ref{fig:lyt} illustrates the layout of the panels presented in the atlas pages. A sample atlas page is provided in Fig. \ref{fig:atlasA963}. The full atlas is available as supplementary material online. The panels in the respective atlas pages are briefly described below.\\

\noindent
(a)\textit{ INT R-band image}: Identified optical counterparts obtained with the INT are encircled in orange. The SDSS ID is indicated in the top-left corner for those galaxies that have SDSS counterparts while some with missing SDSS data have their corresponding INT ID. The optical redshift is given in the bottom-right of the image for those sources with optical spectroscopy. The atlas entries are numbered (top-right) according to their respective index numbers in column 1 of the catalogues. \\

\noindent
(b)\textit{ GALEX NUV image with \H\s contours:} GALEX NUV image of the \H\s source with overlaid \H contours. The contours are set at column density levels of 1, 2, 4, 8, 16 and 32 $\times$ 10$^{19}$ cm$^{-2}$. A red cross indicates the location of the \H centre, while the orange circle indicates the same galaxy as in panel (a). The \H\s ID is provided in the top-left corner.  \\

\noindent
(c) \& (d)\textit{ Position-Velocity diagrams}: The position-velocity slices along the optical major axis were extracted from the R2 (19 km s$^{-1}$) and  R8 (76 km s$^{-1}$) cubes given in the two panels respectively. The red outline indicates the \H mask within which the \H flux was determined. The position angle is given in the top left corner of the diagram. Vertical and horizontal lines correspond to the systemic velocity and the \H\s centre respectively. Contours are drawn at -2 (dashed), 2, 4, 6, 9, 12, 15, 20 and 25 times the local rms noise level.\\

\noindent
(e)\textit{ Global \H profiles}: Global \H profiles for each galaxy correspond to the flux density derived from the R2 cubes (19 km s$^{-1}$), shown in black. The smooth grey line illustrates the \H profiles from cubes smoothed to the R8 resolution. The vertical grey arrow indicates the frequency corresponding to the \H\s redshift of the galaxy, while the orange arrows indicate the frequency corresponding to the optical redshift of the galaxy, if available. Error bars are based on the methodology described in Sect.\ref{gp}.

\section{Summary and Conclusions}\label{summ}
In this paper, we have presented the results of a blind, 21-cm \H\s imaging survey with the WSRT, covering a redshift range of 0.164 < z < 0.224 in each of two pointings, thereby surveying a total volume of 73,400 Mpc$^3$ within the FWQM of the primary beam with a velocity resolution of 19 km s$^{-1}$ and an average angular resolution of 23$\times$38 arcsec$^2$.  We have described in detail the radio data processing procedure, the \H\s source detection methodology and an initial assessment of the completeness of our \H\s detected sample.  The two surveyed volumes each contain an Abell cluster, as well as foreground and background
over-densities.  The \H\s detection limit in the field centres and at the
distances of the clusters is 2$\times$10$^9$ M$_\odot$ over an emission line width of 150 km s$^{-1}$ with a significance of 4$\sigma$ in each of 3 adjacent, independent spectral resolution elements.  A total of 166 galaxies are detected within the total survey volume, none of which have been previously detected in \H\s. We have also presented ancillary optical imaging data of the two fields in the Harris B- and R-bands, obtained with the Wide-Field Camera on the Isaac Newton Telescope on La Palma.  We have described the mosaicing data processing in detail, including the astrometric and photometric calibration.  In addition, we have obtained deep, near- and far-ultraviolet ancillary imaging data of the two fields with the GALEX satellite.  These optical and ultraviolet images have assisted us in identifying the stellar counterparts of the \H\s detected galaxies in case optical redshifts for them were unavailable. We have catalogued the observed and derived \H\s properties of the detected galaxies, such as their \H\s coordinates and redshifts, \H\s line widths, integrated fluxes and \H\s masses.  We have also tabulated the optical and ultraviolet properties of the \H\s detected galaxies such as their SDSS identifiers, available optical spectroscopic redshifts, and their derived absolute R, B, NUV and FUV magnitudes.  The \H, optical and ultraviolet data are compiled in an atlas showing the R-band image from the INT, a contoured \H\s map overlaid on the NUV image from GALEX, position-velocity diagrams at two velocity resolutions, as well as the global \H\s profiles with empirically estimated errors. The \H\s redshift distributions in the two fields outline a well-defined large scale structure along the 328 Mpc deep line-of-sight, with a dominant over-density at the redshifts of the Abell clusters, near-empty voids in front and behind the clusters and smaller over-densities with \H\s detected galaxies in the near-foreground and far-background of those voids.  The volumes surveyed with the WSRT, hence, encompass the widest range of cosmic environments from a deep void to the most dense core of a lensing galaxy cluster.  Abell 963 is a relatively nearby, massive Butcher-Oemler cluster with a large fraction of blue galaxies, while Abell 2192 is much less massive and dynamically less evolved.  The blue galaxies in the core of A963 were not individually detected in \H\s above the survey \H\s mass limit.  Nearly all \H\s detected galaxies associated
with Abell 2192 are located to the South-West of the cluster, indicating an anisotropic accretion of field galaxies.

In a series of forthcoming papers, we will exploit the data provided in this paper to derive the \H\s Mass Function and $\Omega_{HI }$ from direct \H\s detections at z=0.2, study the \H\s-based Tully-Fisher relation at this redshift, and provide an \H\s perspective on the Butcher-Oemler effect. These studies and the \H\s data presented here will also have implications for future \H\s surveys at intermediate redshifts with APERTIF, MeerKAT, ASKAP and the Square Kilometre Array.

\section*{Acknowledgements}
AG and MV acknowledge the Netherlands Foundation for Scientific Research support through VICI grant 016.130.338. We acknowledge the Leids Kerkhoven-Bosscha Fonds (LKBF) for travel support. BD acknowledges the support of the Czech Science Foundation grant 19-18647S and the institutional project RVO 67985815. This research was supported in part by a grant of the US National Science Foundation to Columbia University. JMvdH acknowledges support from the European Research Council under the European Union's Seventh Framework Programme (FP/2007-2013)/ERC Grant Agreement nr. 291531. Y.J. acknowledges financial support from CONICYT PAI (Concurso Nacional de Inserci\'on en la Academia 2017) No. 79170132 and FONDECYT Iniciaci\'on 2018 No. 11180558. AG thanks Kyle A. Oman and Pooja V. Bilimogga for useful comments on the contents of the paper. This work is partly based on observations made with the NASA Galaxy Evolution Explorer (GALEX). The WSRT is operated by the Netherlands Foundation for Research in Astronomy, supported by the Netherlands Foundation for Scientific Research. The full acknowledgement of the Sloan Digital Sky Survey Archive used in this paper can be found at \url{http://www.sdss.org}. IRAF is distributed by the National Optical Astronomy Observatory, which is operated by the Association of Universities for Research in Astronomy (AURA) under cooperative agreement with the National Science Foundation.


\bibliographystyle{mnras}
\bibliography{references}



\bsp	

\appendix
\onecolumn

\setcounter{table}{0}
\renewcommand{\thetable}{A\arabic{table}}

\renewcommand{\thefootnote}{\fnsymbol{footnote}}

\newgeometry{left=4cm,right=4cm,top=2cm}
\onecolumn
\begin{landscape}
\LTcapwidth=\linewidth
\begin{longtable}{ l  l  l  l  l  l  l  l  l  l  l}

\caption{\label{tab:A963_HI_table} A sample table of the \H\s detections in A963. The description of each column is provided in in Sect. \ref{cat}. The full table is available online.} \\

\hline \hline

Index & HI ID  & R.A (J2000) & Dec (J2000) & z$_{\mathrm{HI}}$ & D$_{lum}$ & w$_{20}$ & w$_{50}$ & $S_{int}$ & M$_{HI}$ & Type \\

 &   & hh$:$mm$:$ss.ss & dd$:$mm$:$ss.s &  & Mpc & km s$^{-1}$ & mJy km s$^{-1}$ & mJy km s$^{-1}$ & ($\times$10$^9$) M$_{\odot}$ &  \\
 
(1)  & (2)  & (3) & (4) & (5)  & (6) & (7) & (8) & (9) & (10) & (11)\\

\hline
\endfirsthead
\caption{continued} \\
\hline \hline

Index & HI ID  & R.A (J2000) & Dec (J2000) & z$_{\mathrm{HI}}$ & D$_{lum}$ & w$_{20}$ & w$_{50}$ & $S_{int}$ & M$_{HI}$ & Type \\
 &   & hh$:$mm$:$ss.ss & dd$:$mm$:$ss.s &  & Mpc & km s$^{-1}$ & km s$^{-1}$ & mJy km s$^{-1}$ &($\times$10$^9$) M$_{\odot}$ & \\
(1)  & (2)  & (3) & (4) & (5)  & (6) & (7) & (8) & (9) & (10) & (11) \\

\hline 
\endhead
\hline
\endfoot

1   & HIJ101439.00+385444.8 & 10:14:39.00 & 38:54:44.8 & 0.17361 & 836.7  & 110.2 $\pm$ 9.4  & 83.7 $\pm$ 11.4   & 48.1 $\pm$ 4.2   & 6.77 $\pm$ 0.59  & 2 \\
2   & HIJ101523.89+390949.7 & 10:15:23.89 & 39:09:49.7 & 0.1904  & 926.7  & 349.1 $\pm$ 9.8  & 323.3 $\pm$ 9.5   & 32.4 $\pm$ 1.7   & 5.51 $\pm$ 0.3   & 1 \\
3   & HIJ101535.30+385740.8 & 10:15:35.30 & 38:57:40.8 & 0.21954 & 1086.8 & 222.6 $\pm$ 10.3 & 170.0 $\pm$ 20.8  & 16.2 $\pm$ 1.4   & 3.7 $\pm$ 0.33   & 3 \\
4   & HIJ101536.08+384540.9 & 10:15:36.08 & 38:45:40.9 & 0.20575 & 1010.5 & 222.6 $\pm$ 20.1 & 179.7 $\pm$ 17.0  & 40.5 $\pm$ 2.9   & 8.1 $\pm$ 0.58   & 2 \\
5   & HIJ101536.90+390339.7 & 10:15:36.90 & 39:03:39.7 & 0.2104  & 1036.3 & 206.6 $\pm$ 32.6 & 133.0 $\pm$ 20.4  & 25.8 $\pm$ 1.7   & 5.39 $\pm$ 0.37  & 2 \\
6   & HIJ101540.21+384919.4 & 10:15:40.21 & 38:49:19.4 & 0.20356 & 998.5  & 257.4 $\pm$ 28.6 & 163.4 $\pm$ 44.1  & 46.9 $\pm$ 2.3   & 9.16 $\pm$ 0.46  & 3 \\
7   & HIJ101600.44+385211.7 & 10:16:00.44 & 38:52:11.7 & 0.20861 & 1026.3 & 126.2 $\pm$ 10.4 & 104.3 $\pm$ 12.0  & 18.2 $\pm$ 1.7   & 3.74 $\pm$ 0.36  & 2 \\
8   & HIJ101606.37+385128.0 & 10:16:06.37 & 38:51:28.0 & 0.20739 & 1019.7 & 257.0 $\pm$ 23.1 & 189.3 $\pm$ 29.7  & 37.3 $\pm$ 2.1   & 7.58 $\pm$ 0.43  & 2 \\
9   & HIJ101611.10+384921.7 & 10:16:11.10 & 38:49:21.7 & 0.20674 & 1016.0 & 407.5 $\pm$ 8.1  & 380.3 $\pm$ 10.4  & 66.0 $\pm$ 2.9   & 13.32 $\pm$ 0.59 & 3 \\
10  & HIJ101613.62+390438.4 & 10:16:13.62 & 39:04:38.4 & 0.18954 & 922.2  & 372.1 $\pm$ 15.3 & 315.0 $\pm$ 17.8  & 50.0 $\pm$ 2.0   & 8.44 $\pm$ 0.34  & 3 \\

\end{longtable}

\hspace{2cm}

\begin{longtable}{ l  l  l  l  l  l  l  l  l  l  l }

\caption{\label{tab:A2192_HI_table} A sample table of the \H\s detections in A2192. The description of each column is provided in in Sect. \ref{cat}. The full table is available online.}. \\

\hline \hline
Index & HI ID  & R.A (J2000) & Dec (J2000) & z$_{\mathrm{HI}}$ & D$_{lum}$ & w$_{20}$ & w$_{50}$ & $S_{int}$ & M$_{HI}$ & Type  \\
 &   & hh$:$mm$:$ss.ss & dd$:$mm$:$ss.s &  & Mpc & km s$^{-1}$ & km s$^{-1}$ & mJy km s$^{-1}$ & ($\times$10$^9$) M$_{\odot}$ &  \\
(1)  & (2)  & (3) & (4) & (5)  & (6) & (7) & (8) & (9) & (10) & (11)\\

\hline
\endfirsthead
\caption{continued} \\
\hline \hline

Index & HI ID  & R.A (J2000) & Dec (J2000) & z$_{\mathrm{HI}}$ & D$_{lum}$ & w$_{20}$ & w$_{50}$ & $S_{int}$ & M$_{HI}$ & Type  \\
 &   & hh$:$mm$:$ss.ss & dd$:$mm$:$ss.s &  & Mpc & km s$^{-1}$ & km s$^{-1}$ & mJy km s$^{-1}$ & ($\times$10$^9$) M$_{\odot}$ &  \\
(1)  & (2)  & (3) & (4) & (5)  & (6) & (7) & (8) & (9) & (10) & (11)\\

\hline 
\endhead
\hline
\endfoot

1  & HIJ162451.47+422835.4 & 16:24:51.47 & 42:28:35.4 & 0.18956 & 922.2  & 128.5 $\pm$ 6.5  & 100.2 $\pm$ 11.3 & 27.7 $\pm$ 2.3  & 4.67 $\pm$ 0.38  & 3 \\
2  & HIJ162453.90+422819.1 & 16:24:53.90 & 42:28:19.1 & 0.18915 & 920.2  & 93.6 $\pm$ 5.5   & 70.8 $\pm$ 6.0   & 20.1 $\pm$ 1.6  & 3.38 $\pm$ 0.27  & 2 \\
3  & HIJ162508.17+423409.0 & 16:25:08.17 & 42:34:09.0 & 0.18746 & 910.8  & 154.2 $\pm$ 10.5 & 67.2 $\pm$ 8.5   & 23.7 $\pm$ 1.3  & 3.91 $\pm$ 0.22  & 2 \\
4  & HIJ162510.48+424033.8 & 16:25:10.48 & 42:40:33.8 & 0.18727 & 909.9  & 262.7 $\pm$ 8.5  & 237.9 $\pm$ 12.5 & 35.1 $\pm$ 1.8  & 5.78 $\pm$ 0.3   & 3 \\
5  & HIJ162515.22+423059.6 & 16:25:15.22 & 42:30:59.6 & 0.18656 & 906.1  & 201.8 $\pm$ 9.0  & 167.8 $\pm$ 14.5 & 42.3 $\pm$ 2.0  & 6.91 $\pm$ 0.33  & 2 \\
6  & HIJ162528.35+424708.7 & 16:25:28.35 & 42:47:08.7 & 0.18914 & 919.9  & 140.2 $\pm$ 5.5  & 94.9 $\pm$ 11.0  & 57.2 $\pm$ 2.0  & 9.61 $\pm$ 0.34  & 2 \\
7  & HIJ162533.39+423737.8 & 16:25:33.39 & 42:37:37.8 & 0.20094 & 984.0  & 293.0 $\pm$ 12.6 & 263.2 $\pm$ 17.7 & 26.9 $\pm$ 2.0  & 5.12 $\pm$ 0.38  & 1 \\
8  & HIJ162536.15+425559.3 & 16:25:36.15 & 42:55:59.3 & 0.16842 & 809.0  & 169.7 $\pm$ 19.4 & 140.9 $\pm$ 16.7 & 25.9 $\pm$ 2.4  & 3.42 $\pm$ 0.32  & 2 \\
9  & HIJ162536.33+424132.7 & 16:25:36.33 & 42:41:32.7 & 0.19046 & 927.1  & 84.5 $\pm$ 9.3   & 60.1 $\pm$ 10.5  & 19.0 $\pm$ 1.8  & 3.23 $\pm$ 0.3   & 2 \\
10 & HIJ162544.48+424955.2 & 16:25:44.48 & 42:49:55.2 & 0.17005 & 817.6  & 301.2 $\pm$ 13.3 & 255.2 $\pm$ 20.4 & 69.7 $\pm$ 3.7  & 9.4 $\pm$ 0.5    & 3 \\

\end{longtable}

\end{landscape}
\restoregeometry

\onecolumn
\LTcapwidth=\textwidth
\begin{longtable}{ l  l  l  l  c  l  l  l  l  l }
\caption{\label{tab:A963_optical_table} A sample table containing the optical properties of the \H\s detected galaxies in A963. The description of each column is provided in in Sect. \ref{optcat}. The full table is available online.} \\

\hline \hline

Index & SDSS ID  & R.A (J2000) & Dec (J2000) & P.A & z$_{opt}$ & M$_{B}$ & M$_R$ & M$_{NUV}$ & M$_{FUV}$  \\

 &   & hh$:$mm$:$ss.ss & dd$:$mm$:$ss.s &  $^\circ$  &  &  \\
 
(1)  & (2)  & (3) & (4) & (5)  & (6) & (7) & (8) & (9) & (10)   \\

\hline
\endfirsthead
\caption{continued} \\
\hline \hline

Index & SDSS ID  & R.A (J2000) & Dec (J2000) & P. A & z$_{opt}$ & M$_{B}$ & M$_R$ & M$_{NUV}$ & M$_{FUV}$ \\

 &   & hh$:$mm$:$ss.ss & dd$:$mm$:$ss.s &  $^\circ$  &  &  & & &  \\
 
(1)  & (2)  & (3) & (4) & (5)  & (6) & (7) & (8) & (9) & (10) \\
\hline 
\endhead
\hline
\endfoot

1   & SJ101438.70+385445.0 & 10:14:38.70 & 38:54:45.0 & 2    & 0.17360 & 21.8 & 19.8 & 23.3 & 24.4 \\
2   & SJ101523.65+390941.2 & 10:15:23.65 & 39:09:41.2 & 64   & 0.19030  & 20.0 & 18.5 & 21.5 & 22.2 \\
3   & SJ101534.80+385746.4 & 10:15:34.80 & 38:57:46.4 & 39   & 0.0   & 22.4 & 21.0 &  22.9 & 23.7 \\
4   & SJ101536.22+384532.7 & 10:15:36.22 & 38:45:32.7 & 168  & 0.20555 & 19.8 & 18.6 & 20.6 & 21.2 \\
5   & SJ101537.01+390334.6 & 10:15:37.01 & 39:03:34.6 & 84   & 0.21040  & 21.2 & 19.7 & 22.1 & 22.6 \\
6   & SJ101540.19+384913.7 & 10:15:40.19 & 38:49:13.7 & 89    & 0.20351 & 19.8 & 18.1 & 21.0 & 21.5 \\
7   & SJ101600.10+385205.5 & 10:16:00.10 & 38:52:05.5 & 64    & 0.20866 & 21.2 & 20.1 & 21.5 & 22.0 \\
8   & SJ101605.77+385121.6 & 10:16:05.77 & 38:51:21.6 & 23   & 0.0   & 21.4 & 20.3 & 22.6 & 23.1 \\
9   & SJ101611.13+384924.3 & 10:16:11.13 & 38:49:24.3 & 89    & 0.20600 & 19.2 & 17.4 & 21.0 & 21.6 \\
10  & SJ101613.68+390437.8 & 10:16:13.68 & 39:04:37.8 & 79   & 0.18949 & 20.5 & 18.8 & 22.2 & 22.8 \\
\end{longtable}

\begin{longtable}{ c  c  c  c  c  l  l  l  c  c }

\caption{\label{tab:A2192_optical_table} A sample table containing the optical properties of the \H\s detected galaxies in A2192. The description of each column is provided in in Sect. \ref{optcat}. The full table is available online.} \\

\hline \hline

Index & SDSS ID  & R.A (J2000) & Dec (J2000) & P.A & z$_{opt}$ & M$_{B}$ & M$_R$ & M$_{NUV}$ & M$_{FUV}$ \\

 &   & hh$:$mm$:$ss.ss & dd$:$mm$:$ss.s &  $^\circ$  &  &  &  \\
 
(1)  & (2)  & (3) & (4) & (5)  & (6) & (7) & (8) & (9) & (10) \\

\hline
\endfirsthead
\caption{continued} \\
\hline \hline

Index & SDSS ID  & R.A (J2000) & Dec (J2000) & P. A & z$_{opt}$ & M$_{B}$ & M$_R$ & M$_{NUV}$ & M$_{FUV}$\\

 &   & hh$:$mm$:$ss.ss & dd$:$mm$:$ss.s &  $^\circ$  &  &  &  \\
 
(1)  & (2)  & (3) & (4) & (5)  & (6) & (7) & (8) & (9) & (10) \\
\hline 
\endhead
\hline
\endfoot

1  & SJ162451.64+422828.0 & 16:24:51.64 & 42:28:28.0 & 37  & 0.18943 & 20.9 & 20.0 & 21.3 & 21.7 \\
2  & SJ162454.44+422823.9 & 16:24:54.44 & 42:28:23.9 & 158  & 0.0   & 21.7 & 20.1 & 22.5 & 24.6 \\
3  & SJ162507.59+423408.6 & 16:25:07.59 & 42:34:08.6 & 20  & 0.0   & 19.4 & 18.2 & 20.3 & 21.0 \\
4  & SJ162510.56+424028.8 & 16:25:10.56 & 42:40:28.8 & 111  & 0.18750 & 20.4 & 19.2 & 21.3 & 21.8 \\
5  & SJ162515.35+423057.0 & 16:25:15.35 & 42:30:57.0 & 99 & 0.0   & 20.0\footnote[2]{Galaxies close to bright stars or at the edge of the field (A963 only) with incorrect/unavailable INT magnitudes. The values given are converted from SDSS u, g, r, and i magnitudes.}  & 18.8$^\dagger$  & 19.3 & 19.7 \\
6  & SJ162528.34+424708.8 & 16:25:28.34 & 42:47:08.8 & 4   & 0.18903 & 19.0 & 17.8 & 19.9 & 20.6 \\
7  & SJ162533.32+423742.8 & 16:25:33.32 & 42:37:42.8 & 41  & 0.0   & 22.4$^\dagger$  & 20.8$^\dagger$  & 21.8 & 23.1 \\
8  & SJ162536.21+425558.8 & 16:25:36.21 & 42:55:58.8 & 167  & 0.0   & 20.3 & 19.2 & 21.2 & 21.6 \\
9  & SJ162536.16+424131.8 & 16:25:36.16 & 42:41:31.8 & 31   & 0.19020  & 20.7 & 19.5 & 21.6 & 22.1 \\
10 & SJ162544.36+424953.1 & 16:25:44.36 & 42:49:53.1 & 156  & 0.16797 & 19.5 & 17.6 & 21.4 & 22.0 \\

\end{longtable}







\bsp	
\label{lastpage}

\end{document}